\documentclass[11pt]{article}
\usepackage[preprint]{acl}

\usepackage{times}
\usepackage{latexsym}
\usepackage{float}
\usepackage{paracol}

\usepackage{booktabs}
\usepackage{tablefootnote}
\usepackage{xcolor}
\usepackage{soul}
\usepackage[nameinlink]{cleveref}

\usepackage{enumitem}
\usepackage{makecell}
\usepackage{amssymb} %
\usepackage{pifont}  %
\usepackage{graphicx}
\usepackage{xcolor}   %
\usepackage{annotation}
\usepackage{geometry}
\usepackage[T1]{fontenc}

\definecolor{hlgreen}{HTML}{00e50e}
\definecolor{hlred}{HTML}{ffcccc}

\newcommand{\cmark}{\textcolor{green!70!black}{\checkmark}}
\newcommand{\xmark}{\textcolor{red}{\ding{55}}}

\usepackage[utf8]{inputenc}

\usepackage{microtype}
\usepackage{inconsolata}

\usepackage{graphicx}
\usepackage{multicol}
\usepackage{minted}
\usepackage{listings}

\usepackage{stfloats}

\usepackage[nolist]{acronym}
\usepackage{caption}
\usepackage{subcaption}

\usepackage{colortbl}
\usepackage{xcolor} %
    \definecolor{UBlau}{HTML}{153268}
    \definecolor{LBlau}{HTML}{005f9b}
    \definecolor{LMBlau}{HTML}{0091c8}
    \definecolor{LHBlau}{HTML}{50a5d2}
    \definecolor{WiWi}{HTML}{2b7ab3}
    \definecolor{Grau60}{HTML}{878786}
\usepackage{collcell}

\newcommand{\gradientcell}[3]{%
  \pgfmathsetmacro{\percent}{(#1-#2)/(#3-#2)*100} %
  \edef\tempcellcolor{\noexpand\cellcolor{green!\percent!red!30}}%
  \tempcellcolor #1%
}

\usepackage{adjustbox}
\usepackage{tcolorbox}
\newtcolorbox{combinedprompt}{
    colframe=black,
    colback=white,
    boxrule=0.6mm,
    width=\linewidth,
    fonttitle=\bfseries,
    rounded corners,
    coltitle=black
}

\usepackage{listings} %
\lstnewenvironment{configpython}[2][] %
{
    \lstset{
        commentstyle=\color{UBlau},
        keywordstyle=\color{WiWi},
        numberstyle=\tiny\color{Grau60},
        basicstyle=\ttfamily\footnotesize,
        language=python,
        numbers=none,
        frame=tblr,           %
        tabsize=4,
        captionpos=b,         %
        caption={#1},
        label={#2},
        breaklines=true,      %
        breakatwhitespace=false,
        breakautoindent=true,
        prebreak=\mbox{\textcolor{red}{$\hookrightarrow$}} %
    }
}{}

\lstnewenvironment{configjson}[2][] %
{
    \lstset{
        commentstyle=\color{UBlau},
        keywordstyle=\color{WiWi},
        numberstyle=\tiny\color{Grau60},
        basicstyle=\ttfamily\footnotesize,
        numbers=none,
        frame=tblr,           %
        tabsize=4,
        captionpos=b,         %
        caption={#1},
        label={#2},
        breaklines=true,      %
        breakatwhitespace=false,
        breakautoindent=true,
        prebreak=\mbox{\textcolor{red}{$\hookrightarrow$}} %
    }
}{}

\definecolor{systemcolor}{rgb}{0.9,0.9,1} %
\definecolor{usercolor}{rgb}{1,0.9,0.9} %

\title{MALLM: Multi-Agent Large Language Models Framework}

\author{
  \textbf{Jonas Becker\textsuperscript{1,2,*}, 
  Lars Benedikt Kaesberg\textsuperscript{1,*}, 
  Niklas Bauer\textsuperscript{1},
  Jan Philip Wahle\textsuperscript{1},}\\
  \textbf{Terry Ruas\textsuperscript{1},
  Bela Gipp\textsuperscript{1}} \\
  \textsuperscript{1}University of Göttingen, Germany; \textsuperscript{2}LKA NRW, Germany\\
  $^*$\texttt{\{jonas.becker, l.kaesberg\}@uni-goettingen.de}\\[1em] %
  \begin{tabular}{l@{\hskip 0.5em}l@{\hskip 1em}l}
    \adjustbox{valign=c}{\includegraphics[height=1em]{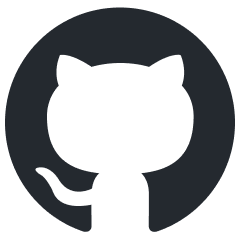}} & Code & \href{https://github.com/Multi-Agent-LLMs/mallm}{github.com/Multi-Agent-LLMs/mallm} \\
    \adjustbox{valign=c}{\includegraphics[height=1em]{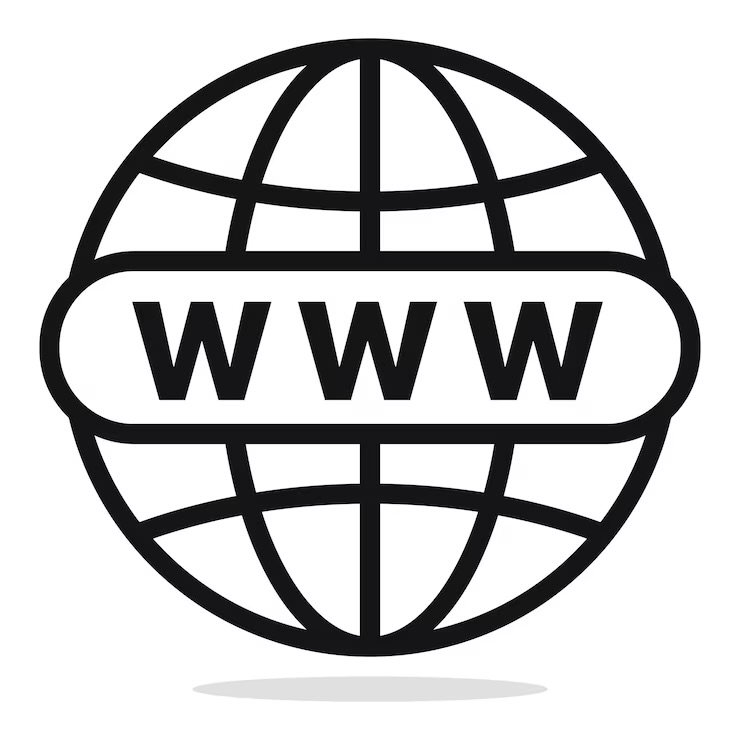}} & Demo & \href{https://mallm.gipplab.org/}{mallm.gipplab.org} \\
    \adjustbox{valign=c}{\includegraphics[height=1em]{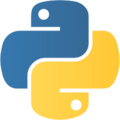}} & Package & \href{https://pypi.org/project/mallm}{pypi.org/project/mallm} %
  \end{tabular}
}

\begin{acronym}
  \acro{AAD}{All-Agents Drafting}
  \acro{CI}{Collective Improvement}
  \acro{CoT}{Chain-of-Thought}
  \acro{LLM}{large language model}
  \acro{MALLM}{\textbf{M}ulti-\textbf{A}gent \textbf{L}arge \textbf{L}anguage \textbf{M}odels}
\end{acronym}

\begin{document}
\maketitle
\AddAnnotationRef{}

\renewcommand{\thefootnote}{\fnsymbol{footnote}}
\footnotetext[1]{Equal contribution.}
\renewcommand*{\thefootnote}{\arabic{footnote}}
\begin{abstract}
Multi-agent debate (MAD) has demonstrated the ability to augment collective intelligence by scaling test-time compute and leveraging expertise.
Current frameworks for MAD are often designed towards tool use, lack integrated evaluation, or provide limited configurability of agent personas, response generators, discussion paradigms, and decision protocols.
We introduce MALLM (\ul{M}ulti-\ul{A}gent \ul{L}arge \ul{L}anguage \ul{M}odels), an open-source framework that enables systematic analysis of MAD components.
MALLM offers more than 144 unique configurations of MAD, including (1) agent personas (e.g., \texttt{Expert}, \texttt{Personality}), (2) response generators (e.g., \texttt{Critical}, \texttt{Reasoning}), (3) discussion paradigms (e.g., \texttt{Memory}, \texttt{Relay}), and (4) decision protocols (e.g., \texttt{Voting}, \texttt{Consensus}).
MALLM uses simple configuration files to define a debate.
Furthermore, MALLM can load any textual Hugging Face dataset (e.g., MMLU-Pro, WinoGrande) and provides an evaluation pipeline for easy comparison of MAD configurations.
MALLM enables researchers to systematically configure, run, and evaluate debates for their problems, facilitating the understanding of the components and their interplay.
\end{abstract}

\section{Introduction}
\label{sec:introduction}

\begin{figure*}
    \centering
    \includegraphics[width=\linewidth]{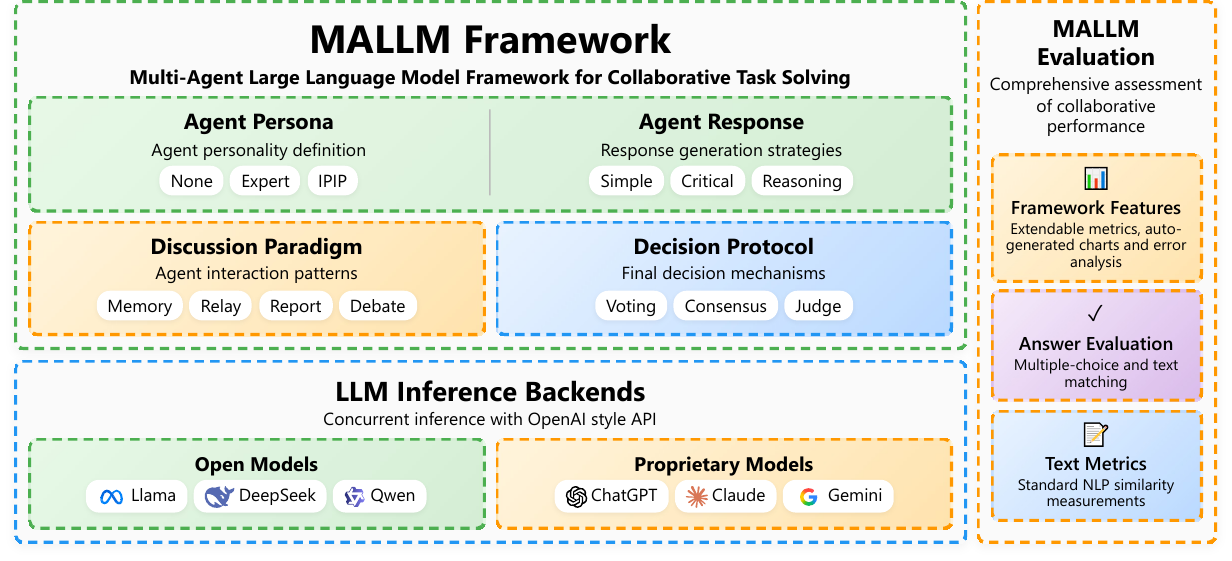}
    \caption{Overview of the MALLM framework and its components.
    }
    \label{fig:mallm-framewrok}
\end{figure*}

Multi-agent debate (MAD) has emerged as a new paradigm to solve complex tasks with multiple large language models (LLMs) \cite{Chan2023ChatEvalTBA, DuLTT23a, Liang2023EncouragingDTA, wang_rethinking_2024}.
Yet, we have not understood the exact mechanisms of when and why MAD is successful.
Different hypotheses exist around whether MAD is another way to scale test-time compute \cite{Yang2025RevisitingMDA}, or whether the combination of individual components has emergent capabilities \cite{Liang2023EncouragingDTA}.
Understanding these mechanisms requires a systematic evaluation, specifically code that enables adjusting one variable of the MAD at a time to measure its effect.

Recent work \citep{GuoCWC24, tran2025multiagentcollaborationmechanismssurvey, tillmann2025literaturereviewmultiagentdebate} has identified several key aspects influencing multi-agent discussions. 
We focus on the following three main components: (1) \textbf{agents} define ``who'' is participating in the debate, meaning the personas of agents and their response style \citep{WangMWG23,XuYLW23}; (2) \textbf{discussion paradigms} determine ``how'' the debate is taking place, including agent response order and turn boundaries, structuring information flow \citep{YinSCG23a}; (3) \textbf{decision protocols} choose ``what'' the debate result will be, meaning deciding when discussions end and determining a final answer \citep{chen2025multiagentconsensusseekinglarge,kaesberg2025votingconsensusdecisionmakingmultiagent}.
As later evaluations will demonstrate, each of these components is crucial in downstream tasks.
Adjusting them individually is particularly important for systematic investigations.

To satisfy the growing demand for MAD applications, many frameworks have been developed \citep{wu_autogen_2023, GaoLPK24, hong_metagpt_2024}.
Yet these frameworks intertwine the definitions of multiple components, such as agents and discussion paradigms \citep{wu_autogen_2023} or do not allow for adjusting specific parts, such as discussion paradigms or decision protocols \citep{hong_metagpt_2024}, hindering independent experimentation.
Existing approaches often constrain experimentation by using predefined agents, discussion paradigms, or decision protocols \citep{zhuge_language_2024, GaoLPK24}. 
Some frameworks also specialize in particular use cases, such as tool usage \citep{openai_agents_python}, and typically lack integrated evaluation pipelines for analyzing individual MAD components systematically \citep{wu_autogen_2023, GaoLPK24}. 
To the best of our knowledge, no current framework exists to evaluate the three core components of related works and their interactions for MAD: agents, discussion paradigms, and decision protocols.

We propose the open-source \ul{M}ulti-\ul{A}gent \ul{LLM} (MALLM) framework to address these limitations (see \Cref{fig:mallm-framewrok} for an overview).
MALLM integrates several MAD components from previous research.
Researchers can individually configure their debate setup through parameter settings and easily extend the framework by inheriting existing classes or using template functions.
Without additional programming effort, MALLM supports more than 144 distinct MAD configurations.
Thus, it enables researchers to reproduce prior MAD experiments, apply MAD methods to new datasets or tasks, and systematically ablate individual MAD components to analyze their impact.
We present MALLM's capabilities on our demo website\footnote{\href{https://mallm.gipplab.org/}{mallm.gipplab.org}}.

\noindent\textbf{The key contributions are:}
\begin{itemize}
    \item We propose MALLM, an open-source framework that supports studies of MAD by enabling controlled variation of agents, discussion paradigms, and decision protocols.
    \item We allow researchers to experiment with 144 existing MAD configurations on a wide range of text-based tasks through automated dataset loading, preprocessing, and evaluation. %
    \item We provide abstract classes and template functions to implement new MAD components and tasks, allowing others to understand the best configuration for their specific research goals.
\end{itemize}

\section{Related Work}
\label{sec:related_work}

We identify three key components of related work that are commonly discussed in MAD.
\citet{WangMWG23, XuYLW23} use \textbf{agents} with varying personas and response styles.
\citet{YinSCG23a, li2024improvingmultiagentdebatesparse} define \textbf{discussion paradigms} that determine turn order and information flow.
\citet{chen2025multiagentconsensusseekinglarge, YangDKH24} explain variations in \textbf{decision protocols} that aggregate agent outputs into a single solution.

Existing MAD frameworks vary across the core components described previously: agents, discussion paradigms, and decision protocols. 
AutoGen supports multi-turn interactions between customizable agents but does not separate discussion paradigms from agent definitions, hindering systematic studies of each component \citep{wu_autogen_2023}. 
It also does not support integrated evaluation pipelines, which need to be coded externally. 
Similarly, MetaGPT assigns tasks to specialized agents that follow standard operating procedures, tightly coupling agent roles and response styles, but restricting experimentation with alternative discussion paradigms or decision protocols \citep{hong_metagpt_2024}.

Other frameworks offer alternative abstractions. 
GPTSwarm models agent interactions as optimizable computational graphs, focusing on information flow rather than the modular comparison of agents or decision protocols \citep{zhuge_language_2024}. 
AgentScope simplifies interactions using predefined pipelines \citep{GaoLPK24}, constraining discussion paradigms and limiting evaluation of agent personas or decision protocols. 
The OpenAI Agents SDK coordinates tool-using agents \citep{openai_agents_python}, prioritizing agent functionality but lacking customizable decision protocols for MAD.

A common limitation across existing frameworks is the tight coupling among agents, discussion paradigms, and decision protocols, which hinders the analysis of each component independently or in combinations of choice. 
This makes the investigation of which specific components contribute and should be used for particular use cases difficult.
Most frameworks provide fixed orchestration setups, restricting experimentation with alternative decision protocols or agent configurations \citep{openai_agents_python, zhuge_language_2024}, and few include integrated evaluation pipelines \citep{wu_autogen_2023, GaoLPK24, smit2024goingmadlookmultiagent}. 
No current framework explicitly supports the systematic analysis of individual MAD components and their interactions. 
Our proposed framework, MALLM, addresses these limitations with a modular architecture clearly separating agents, discussion paradigms, and decision protocols into interchangeable modules. 
This design enables a systematic study of each component independently and combined, supported by an integrated evaluation pipeline.
A comparison of MALLM and other frameworks for MAD is included in \Cref{tab:frameworks} of \Cref{app:other_frameworks}.

\section{MALLM Framework}
\label{section:framework}

We propose MALLM, a framework for MAD. 
It coordinates agents to solve text-based tasks \citep{GuoCWC24}.
MALLM receives an input task and outputs a solution after performing a MAD. 
\Cref{fig:mallm-framewrok} illustrates the components of MALLM.

MALLM implements three \textbf{agent personas} (\texttt{None}, \texttt{Expert}, \texttt{IPIP}), three \textbf{agent response generators} (\texttt{Simple}, \texttt{Critical}, \texttt{Reasoning}), four \textbf{discussion paradigms} (\texttt{Memory}, \texttt{Relay}, \texttt{Report}, \texttt{Debate}), and three main \textbf{decision protocols} (\texttt{Voting}, \texttt{Consensus}, \texttt{Judge}).
Each of the component variants can be parameterized individually, allowing systematic comparison of individual setups without additional code.

The agents participating in the debate can use most proprietary and open models, as MALLM supports any OpenAI-compatible API endpoint for inference.
The integrated evaluation pipeline can be used to analyze the large amounts of data generated by MAD in a unified way and directly generate comparative charts to visualize performance across different configurations.
We provide more details on the framework parameters in \Cref{app:parameters} and prompts in \Cref{app:prompts}.

With MALLM, users can explore the effects of changing components within MAD.
The effects of each variation in MAD can be measured towards solving various text-based problems, such as mathematical reasoning \citep{cobbe2021trainingverifierssolvemath}, ethical question-answering \citep{hendrycks2023aligningaisharedhuman}, and more.
Our public demo includes three persona generators, three response generators, four discussion paradigms, and four decision protocols\footnote{\href{https://mallm.gipplab.org/}{mallm.gipplab.org}}.
Thus, users can explore 144 MAD configurations directly, observing the effect of parameter combinations.
A screenshot of our interactive demo is in \Cref{fig:app_mallm_screenshot} of \Cref{app:demo}.

\subsection{Agents}

An \textbf{agent} is defined by its role (persona generator) and its answer style (response generator). 

\paragraph{Persona Generator.}

The persona specifies agent behavior by their system prompt, e.g., expertise or personality \citep{XuYLW23, WangMWG23}. %
Personas are created iteratively to be complementary and unique. 
We include three persona types: \texttt{None}, \texttt{Expert}, and \texttt{IPIP}. 

\noindent\textbf{None}: Disables the persona generation. It assigns each agent a generic name (``Participant N'') for baseline experiments.
\noindent\textbf{Expert}: Creates domain-specific personas aligned with the task description \citep{XuYLW23}. Examples are an ``Educator'' for machine learning explanations, a ``Software Developer'' for app development, or a ``Chef'' for cooking tasks.
\noindent\textbf{IPIP}: Based on the Big Five personality traits, using the open-source IPIP-NEO classification \citep{costa1992normal,goldberg1999broad}. They cover Extraversion, Agreeableness, Conscientiousness, Neuroticism, and Openness. The items originate from the International Personality Item Pool (IPIP), frequently used in Psychology \citep{maples2014test}. This enables the detailed modeling of psychological diversity \citep{serapio-garcia_personality_2023, sorokovikova2024llms}.

\paragraph{Response Generator.}

The response generator produces agent responses in a specified format or style, influencing how agents interact (e.g., neutral or critical) \citep{10.1162/tacl_a_00681}.
They vary in how the agents are prompted to generate a response in the debate.
MALLM includes the \texttt{Simple}, \texttt{Reasoning}, and \texttt{Critical} response generators.
We include their prompts in \Cref{app:prompts_responsegenerator}.

\noindent\textbf{Simple}: Produces free-text responses in a neutral tone, explicitly indicating agreement or disagreement.
\noindent\textbf{Reasoning}: Responds step-by-step, including analysis, alternatives, and conclusions. Agents share their reasoning but not solutions, encouraging independent idea generation. 
\noindent\textbf{Critical}: Tasks the agent to identify weaknesses, question assumptions, and suggest alternative approaches.

\subsection{Discussion Paradigm}

\begin{figure}[t]
    \centering
    \includegraphics[width=\linewidth]{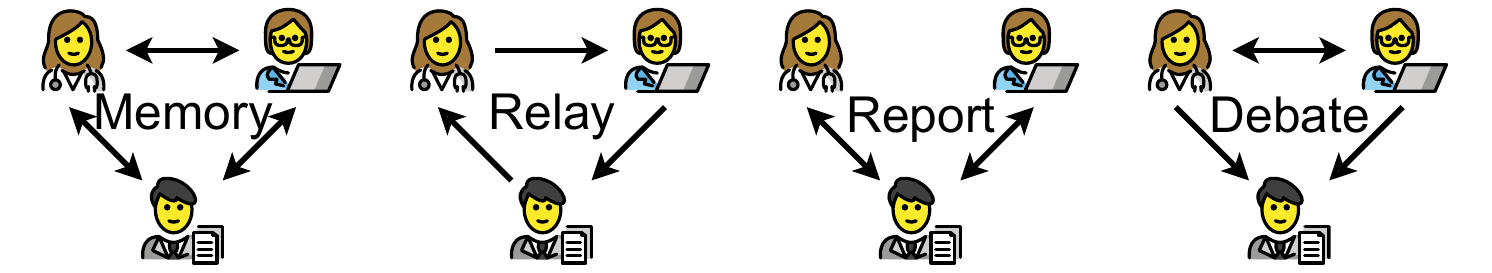}
    \caption{Overview of four discussion paradigms.}
    \label{fig:mallm-paradigms}
\end{figure}

The \textbf{discussion paradigm} defines the structure of agent interaction. 
It specifies turn-taking and information access rules for the MAD. 
We implement four paradigms \citep{YinSCG23a}: \texttt{Memory}, \texttt{Report}, \texttt{Relay}, and \texttt{Debate}.
Each paradigm differs in information flow and visibility, as illustrated in \Cref{fig:mallm-paradigms}.

\noindent\textbf{Memory:} All agents have full visibility into each other's messages across turns.
\noindent\textbf{Relay:} Information is passed sequentially between agents in a chain, with only the last message visible to the next.
\noindent\textbf{Report:} Agents independently solve the tasks and report back to a central agent.
\noindent\textbf{Debate:} Agents argue in pairs, taking turns to debate intermediate conclusions before a central agent is consulted.

\subsection{Decision Protocol}

Each agent in a multi-agent system generates solution drafts.
\textbf{Decision protocols} systematically determine when discussions end and combine agent-generated solutions into a final decision.

\begin{figure}[t]
    \centering
    \includegraphics[width=\linewidth]{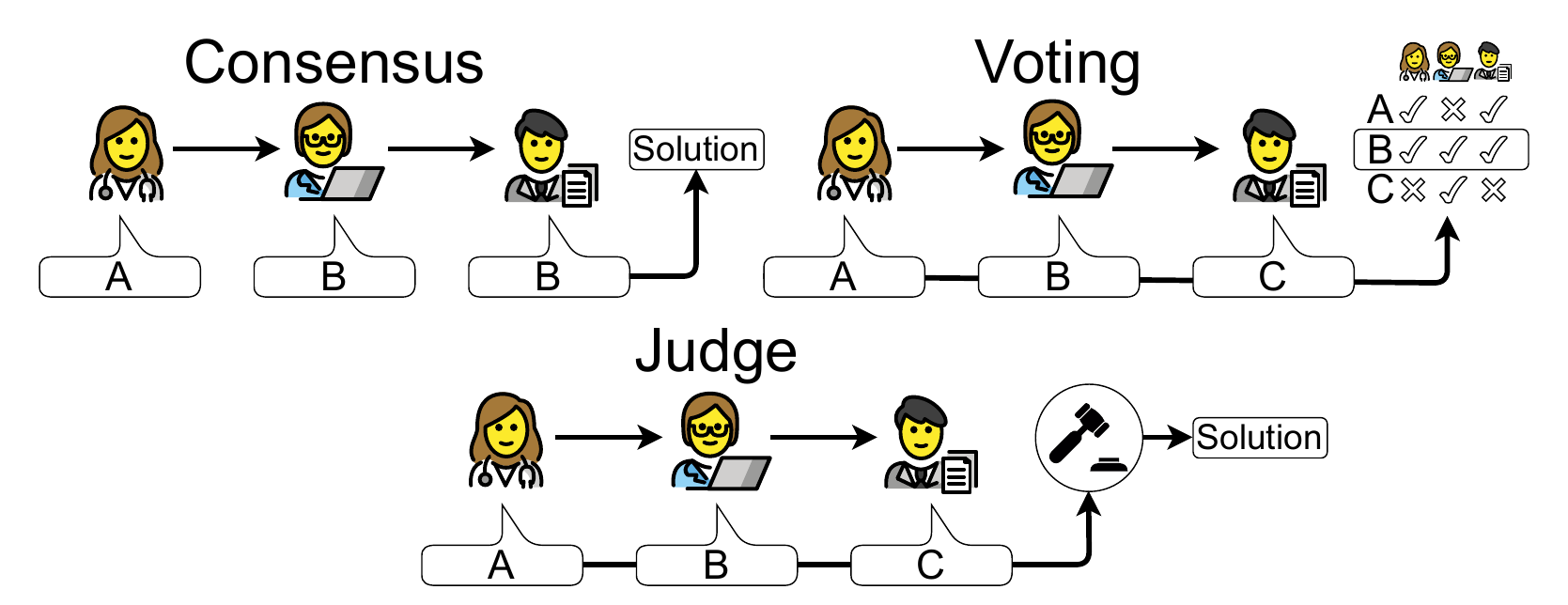}
    \caption{Overview of three main decision protocols.
    }
    \label{fig:mallm-decision}
\end{figure}

The MALLM framework implements three decision protocol families: \texttt{Consensus}, \texttt{Voting}, and \texttt{Judge}, as illustrated in \Cref{fig:mallm-decision}.
Debates can run with a fixed number of turns or perform early stopping upon a successful decision.

\noindent\textbf{Consensus}: Consensus decision protocols decide on an answer by having the agents converge on one solution. The solution is selected when a required level of agreement among agents is reached \citep{YinSCG23a}. MALLM includes three agreement levels: \texttt{Majority Consensus} (over 50\%), \texttt{Supermajority Consensus} (over 66\%), and \texttt{Unanimity Consensus} (100\%).
\noindent\textbf{Voting}: Uses a fixed number of discussion rounds (default three, following findings by \citet{DuLTT23a}) before agents vote. In the event of a tie, we run an additional round of debate and voting. Variants are inspired by \citet{YangDKH24} and include \texttt{Simple Voting} (i.e., each agent votes for their preferred solution), \texttt{Approval Voting} (i.e., multiple acceptable solutions per agent), \texttt{Ranked Voting} (i.e., agents rank solutions, best cumulative rank selected), and \texttt{Cumulative Voting} (i.e., agents allocate up to 25 points across solutions, highest total points selected).
\noindent\textbf{Judge}: Relies on one agent reviewing all solutions, choosing either a preferred one or synthesizing a new solution. The effectiveness of the Judge protocol depends on the model's reasoning capabilities \citep{zheng_judging_2023}.

\subsection{Evaluation}

The MALLM framework includes a pipeline for evaluating MAD configurations, producing statistics and charts for a dataset and its metrics.

\noindent\textbf{Datasets.} The pipeline provides integrated loaders for reasoning tasks (e.g., WinoGrande \citep{sakaguchi2019winograndeadversarialwinogradschema}, StrategyQA \citep{GevaKSK21}) and knowledge tasks (e.g., GPQA \citep{rein_gpqa_2023}, MMLU-Pro \citep{WangMZN24}) as well as core tasks of text generation \citep{becker2024textgenerationsystematicliterature}, such as paraphrasing \citep{kovatchev-etal-2018-etpc} or summarization \citep{narayan-etal-2018-dont}.
It further supports any textual Hugging Face dataset for problem-solving.
Researchers can also add their own dataset via subclassing.
All datasets are converted into a unified format for processing.

\noindent\textbf{Metrics.} We include question-answering and free-text evaluations.
For question-answering, we compute accuracy by comparing selected response letters against reference solutions via regex.
For free-text tasks, we include BERTScore \citep{zhang2020bertscoreevaluatingtextgeneration} and textual overlap measures (BLEU \citep{PapineniRWZ02a}, ROUGE-1/2/3/L \citep{lin-2004-rouge}, and METEOR \citep{banerjee-lavie-2005-meteor}).

\noindent\textbf{Statistical variance.} 
We find that several studies on MAD do not account for statistical variance \citep{wu_autogen_2023,talebirad2023multiagentcollaborationharnessingpower}, yet it can have marked impacts on MAD \citep{smit2024goingmadlookmultiagent}.
MALLM enables repeated experiments and calculates standard deviations between them, thereby quantifying statistical dispersion.

\noindent\textbf{Automatic charts.} MALLM can visualize the evaluation results of MAD configurations.
For this, researchers can pass a single file or a directory with multiple evaluation results directly to an automatic chart generator.
Examples of the generated charts by the evaluation pipeline can be seen in \Cref{fig:mallm-auto-clock,fig:mallm-auto-score,fig:mallm-auto-decision,fig:mallm-auto-turns} of \Cref{app:auto-charts}.

\section{Application}

\begin{figure*}[ht!]
    \centering
    \includegraphics[width=\linewidth]{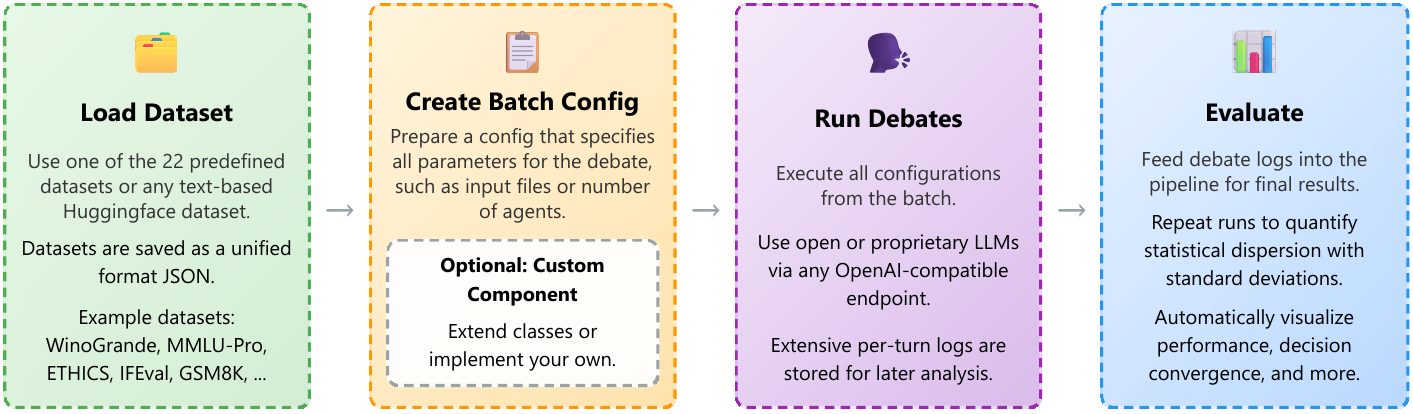}
    \caption{Example workflow of experimenting with MAD using MALLM. First, we can load a dataset. Second, a config file defines the MAD. Third, the debates run and produce output logs. Last, the debates are evaluated. While MALLM already comes with many parameters and components to test, researchers can optionally incorporate their own components, which are tailored to their specific experiment.}
    \label{fig:app_mallm_workflow}
\end{figure*}

We explain the workflow for using MALLM to evaluate a specific MAD configuration. 
First, we can load any text-based Hugging Face dataset for problem-solving via our dataset loader (e.g., MMLU-Pro \citep{WangMZN24}).
Second, MALLM comes with a configuration file (cf. \Cref{app:batch-config-file}) that the user can change to adjust parameters and the components for the desired experiment (e.g., using \texttt{Llama-3.3-70B-Instruct} as a model and changing the configuration to \texttt{four} debating agents, \texttt{Relay} discussion paradigm, and \texttt{Unanimity Consensus} decision protocol). 
Third, we can run the experiment specified by the configuration file. 
MALLM keeps extensive logs of all debates.
They include each agent's messages, votes cast, used components, and configuration parameters.
Lastly, we can directly feed the logs into our evaluation pipeline, computing metrics (e.g., accuracy) and leveraging automatic chart generation (cf. \Cref{fig:mallm-auto-clock,fig:mallm-auto-score,fig:mallm-auto-turns,fig:mallm-auto-decision} in \Cref{app:auto-charts}).

Each component (agent, discussion, decision) comes with abstract base classes to describe the pipeline of MAD.
If necessary, a custom component could be provided by inheriting features from a component's abstract base class.
For example, Sketch-of-Thought \citep{aytes2025sketchofthoughtefficientllmreasoning} is a variant of \acl{CoT} \citep{WeiWSB23} that restructures how models express intermediate steps of their reasoning.
To integrate with MALLM, we would create a custom class \texttt{SoTResponseGenerator} inheriting from the abstract base class \texttt{ResponseGenerator}. 
Then, the components of MAD can be defined by a configuration file as usual.
\Cref{fig:app_mallm_workflow} provides an overview of the workflow for conducting and evaluating MAD with MALLM.

\subsection{Use Cases}

The following examples illustrate possible research directions that can be realized using MALLM.

\noindent\textbf{Agents on paradigms.} The number of agents for the MAD is modifiable. Researchers can test the impact of this on the various discussion paradigms. For example, a study could compare two, three, four, or five agents on the memory and relay paradigm. As these paradigms differ in their information flow, it would be interesting to see how this impacts task performance.

\noindent\textbf{Testing new task.} JailbreakBench \citep{chao2024jailbreakbenchopenrobustnessbenchmark} measures the safety of LLMs against jailbreaks. One avenue could be the comparison of multi-agent safety with a single-agent setup. For this, we can use MALLM, which is plug-and-play with a template configuration set.

\noindent\textbf{Finetuned agents.} MALLM works with any proprietary and open-source LLM, which means that we can also provide our own model for the agents. A promising direction is to finetune an agent to enhance its argumentation skills, thereby improving task performance on reasoning tasks such as StrategyQA \citep{GevaKSK21}.

\noindent\textbf{Moderated paradigm.} A dynamic moderator can be implemented by subclassing the abstract base class \texttt{DiscussionParadigm}. We can define logic for an LLM-based moderator agent to adjust speaking order based on previous agent contributions. This enables an investigation into the effects of adaptive moderation.

\subsection{Example Experiments with MALLM}
MALLM can be applied to various use cases.
To demonstrate the opportunities for experimental setups, we provide some example investigations. 
Supplementary information, such as used models and parameters, can be found in \Cref{app:parameters}.

\begin{table}[t!]
\small
\centering

\begin{tabular}{ccc}
\toprule
\textbf{Simple} & \textbf{Critical} & \textbf{Reasoning} \\
\midrule
58.6\tiny{$\pm$1.6} & \textbf{61.4\tiny{$\pm$3.3}} & 52.2\tiny{$\pm$2.8} \\
\bottomrule
\end{tabular}

\caption{Comparison of accuracy averaged over all voting-based decision protocols using the simple, critical, and reasoning response generators on the StrategyQA dataset. Best is bolded. $\pm$ shows standard deviation over three runs.}
\label{tab:response_gen}
\end{table}

\begin{table}[t]
\centering
\small
\begin{tabular}{ccccc}
\toprule
\textbf{CoT} & \textbf{Memory} & \textbf{Relay} & \textbf{Report} & \textbf{Debate} \\
\midrule
\textbf{56.9}\tiny{$\pm$1.8} & 60.8\tiny{$\pm$2.6} & \textbf{62.9\tiny{$\pm$1.6}} & 60.9\tiny{$\pm$3.1} & 61.9\tiny{$\pm$1.1} \\
\bottomrule
\end{tabular}
\caption{Accuracy of MAD on StrategyQA with different discussion paradigms. Best and worst are bolded. $\pm$ shows standard deviation over five runs.
}
\label{table:performance_paradigms}
\end{table}

\noindent\textbf{Agents.} \citet{kaesberg2025votingconsensusdecisionmakingmultiagent} use MALLM to experiment with response generators on the StrategyQA dataset \citep{GevaKSK21}.
\Cref{tab:response_gen} presents the average accuracy across all voting-based decision protocols, using the \texttt{Memory} discussion paradigm and \texttt{Expert} personas with the \texttt{Simple}, \texttt{Critical}, and \texttt{Reasoning} response generators.
The \texttt{Critical} response generator slightly improves performance by encouraging agents to critically evaluate responses from others, resulting in a 2.8\% point increase.
The \texttt{Reasoning} response generator decreases performance by 6.4\% points, likely because it imposes a strict response structure.
Strictly structured responses can degrade the task performance of MAD, a characteristic that was previously noted in single-agent setups \citep{tam2024letspeakfreelystudy}.
To summarize, invoking strict response patterns from agents can harm task performance, while prompting agents to think critically can boost it.

\begin{table}[t]
\centering
\small
\begin{tabular}{lcc}
\toprule
\textbf{Dataset} & \textbf{Voting} & \textbf{Consensus} \\ \midrule
\textbf{Knowledge-Based}\\
 \quad MMLU       & 51.7 \tiny{$\pm$ 2.4} & \textbf{54.0 \tiny{$\pm$ 2.7}} \\
 \quad MMLU-Pro   & 31.1 \tiny{$\pm$ 3.5} & \textbf{36.0 \tiny{$\pm$ 1.8}} \\
 \quad GPQA       & 29.7 \tiny{$\pm$ 2.5} & \textbf{31.0 \tiny{$\pm$ 2.4}} \\ \midrule
\textbf{Reasoning-Based}\\
 \quad SQuAD 2.0     & \textbf{56.7 \tiny{$\pm$ 1.6}} & 43.6 \tiny{$\pm$ 1.5} \\
 \quad StrategyQA & \textbf{58.6 \tiny{$\pm$ 2.0}} & 58.4 \tiny{$\pm$ 1.6} \\
 \quad MuSR       & \textbf{54.8 \tiny{$\pm$ 1.9}} & 28.4 \tiny{$\pm$ 2.6} \\
 \bottomrule
\end{tabular}
\caption{Mean performance for voting and consensus decision protocols on knowledge and reasoning tasks. Best is bolded. $\pm$ shows standard deviation over three runs. %
}
\label{tab:dataset_vote_cons_std}
\end{table}

\noindent\textbf{Discussion Paradigms.} \citet{becker_multi-agent_2024} compares discussion paradigms on the StrategyQA dataset \citep{GevaKSK21}, using \texttt{Expert} personas, the \texttt{Simple} response generator, and \texttt{Majority Consensus}.
MAD runs until the agent agrees to a solution or until a maximum of seven turns is reached.
We find that, using \texttt{Majority Consensus}, most debates reach an agreement and end within the first three turns.
Thus, seven turns provide a reasonable headroom for this experiment.
The results reveal two findings.
First, \Cref{table:performance_paradigms} compares the discussion paradigms of MAD against a single LLM baseline with \acl{CoT} \citep{WeiWSB23}.
All paradigms outperform a single LLM with \acl{CoT} prompting on StrategyQA, improving accuracy by up to 4.0\%.
Second, we investigate the impact of information transparency on convergence speed for MAD.
We find that the very transparent paradigm \texttt{Memory} enables faster consensus (avg. 1.75 turns), while limited visibility between agents in \texttt{Relay} slows it down (avg. 2.61 turns).
Thus, information transparency can lead to quicker convergence in MAD without sacrificing task performance.
In summary, MAD can outperform Chain-of-Thought on reasoning tasks, such as StrategyQA, while the information transparency of the discussion paradigm impacts the convergence speed of MAD.

\noindent\textbf{Decision Protocols.} Investigations by \citet{kaesberg2025votingconsensusdecisionmakingmultiagent} compare the average of all \texttt{Voting} and \texttt{Consensus} decision protocols across knowledge and reasoning tasks, summarized in \Cref{tab:dataset_vote_cons_std}. They use the \texttt{Simple} response generator and the \texttt{Memory} discussion paradigm. \texttt{Consensus} consistently outperforms \texttt{Voting} on knowledge tasks (MMLU \citep{hendrycks_measuring_2021-1}, MMLU-Pro \citep{WangMZN24}, GPQA \citep{rein_gpqa_2023}), achieving approximately 2.8\% higher accuracy due to repeated verification steps. \texttt{Voting} protocols significantly improve accuracy by about 13.2\% on reasoning-intensive tasks (SQuAD 2.0 \citep{rajpurkar_know_2018}, StrategyQA \citep{GevaKSK21}, MuSR \citep{sprague2024musrtestinglimitschainofthought}), benefiting from diverse reasoning paths. To summarize, the selection of the decision protocol depends on the specific task. When chosen correctly, it can notably improve task performance.

\noindent\textbf{Creating Demo Examples.} For the demonstration of MALLM, we create a set of 144 different example configurations using MALLM's batch feature, called DEBATE (\ul{D}iverse \ul{E}xchanges \ul{B}etween \ul{A}utonomous \ul{T}alking \ul{E}ntities).
We release the DEBATE dataset publicly\footnote{\href{https://huggingface.co/datasets/Multi-Agent-LLMs/DEBATE/}{huggingface.co/datasets/Multi-Agent-LLMs/DEBATE}}.
Potential uses for this data could be to (1) study how agents debate as a proxy to humans; (2) study the structural reasons why MAD can fail in some scenarios, as identified by \citet{becker2025stayfocusedproblemdrift}; (3) explore how prompting agents to assess prior messages critically affects the speed of consensus-building.
More details on the DEBATE dataset are in \Cref{app:dataset}.

\section{Epilogue}
\label{sec:epilogue}

We proposed MALLM, a framework specialized in conversational problem-solving for MAD. 
MALLM enables users to configure debates for their specific problems and research objectives.
More specifically, our framework supports the analysis of multiple components, including agent personas, response generators, discussion paradigms, and decision protocols. 

MALLM works with most proprietary and open models and can load any text-based Hugging Face dataset for problem-solving.
MALLM's evaluation pipeline offers pre-implemented metrics (e.g., Accuracy, BLEU, BERTScore) and automatic chart generation, while accounting for the statistical variance of MAD.
A demo for the capabilities of MALLM is publicly available\footnote{\href{https://mallm.gipplab.org/}{mallm.gipplab.org}}.

We described four potential use cases that can be further developed: the number of agents and their impact, the resilience of MAD against jailbreak attacks, finetuning specialized agents, and adaptive moderation.
Future work could also expand MALLM with additional functionalities, such as evaluating debates through any Hugging Face metric.
Beyond serving as a testbed for MAD itself, MALLM provides researchers with an environment to assess how variations in agents, discussion paradigms, and decision protocols can impact their specific problems.

\section*{Limitations}
\label{sec:limitations}

We provide the MALLM framework with pre-implemented variants  for personas, response generators, discussion paradigms, and decision protocols.
While our goal was to include diverse variants backed by literature (e.g., voting \citep{YangDKH24}, consensus \citep{YinSCG23a}, and judge \citep{zheng_judging_2023} for decision protocols), we could not account for all possible patterns of agent orchestration.
There is the possibility that niche use cases with MAD and future developments are not captured by our selection, e.g., decisions by confidence-weighted voting \citep{chen2024reconcileroundtableconferenceimproves}.
We publish our framework as open-source, allowing and encouraging researchers to develop custom components via subclassing and apply MALLM to their specific use cases.

\section{Acknowledgements}
This work was supported by the Landeskriminalamt NRW, the Lower Saxony
Ministry of Science and Culture and the VW Foundation.
We acknowledge EuroHPC Joint Undertaking for awarding us access to MeluXina at LuxProvide, Luxembourg.
This work was funded by the Deutsche Forschungsgemeinschaft (DFG, German Research Foundation) – 564661959.

\bibliography{merged_deduplicated}

\appendix

\clearpage
\section*{\Large Appendix}
\section{MALLM Workflow}
\label{app:workflow}
\begin{figure*}[!b]
    \centering
    \includegraphics[width=\linewidth]{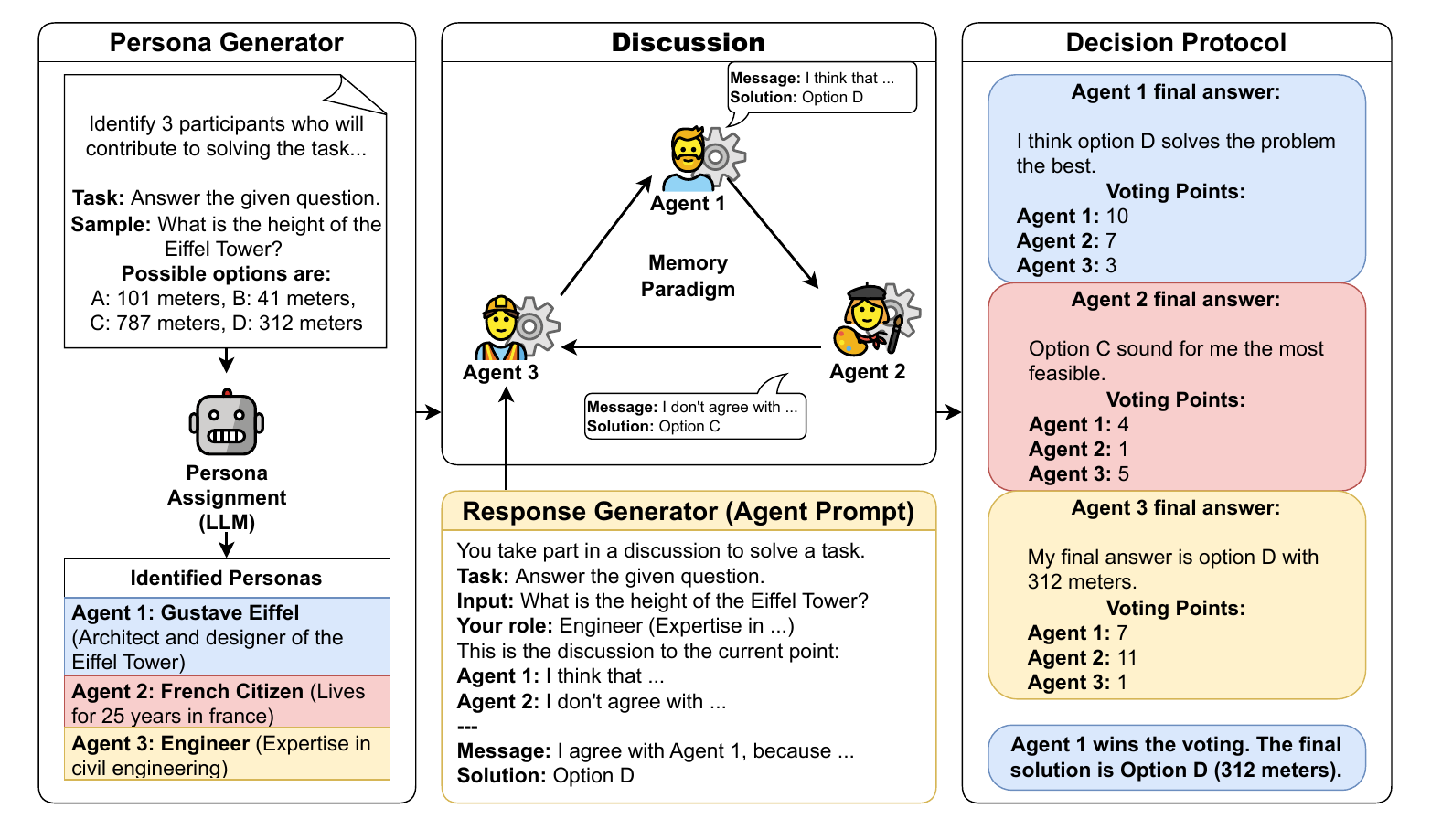}
    \caption{Example multi-agent discussion conducted in the MALLM framework. It showcases the functionality of the four modules and how they work together to get an improved final solution. First, we use the  \texttt{Expert} persona generator to create three agents with different expertise. These agents discuss according to the \texttt{Memory} discussion paradigm and use the \texttt{Simple} response generator to formulate their answers. After the third turn, they begin voting using the \texttt{Cumulative Voting} decision protocol until they reach a final solution.}
    \label{fig:app_mallm_overview}
\end{figure*}

We provide an example of how MALLM could be used to conduct experiments on MAD in \Cref{fig:app_mallm_workflow}. 
An example discussion can be seen in \Cref{fig:app_mallm_overview}.

\section{System Demonstration} \label{app:demo}

\begin{figure*}
    \centering
    \includegraphics[width=\linewidth]{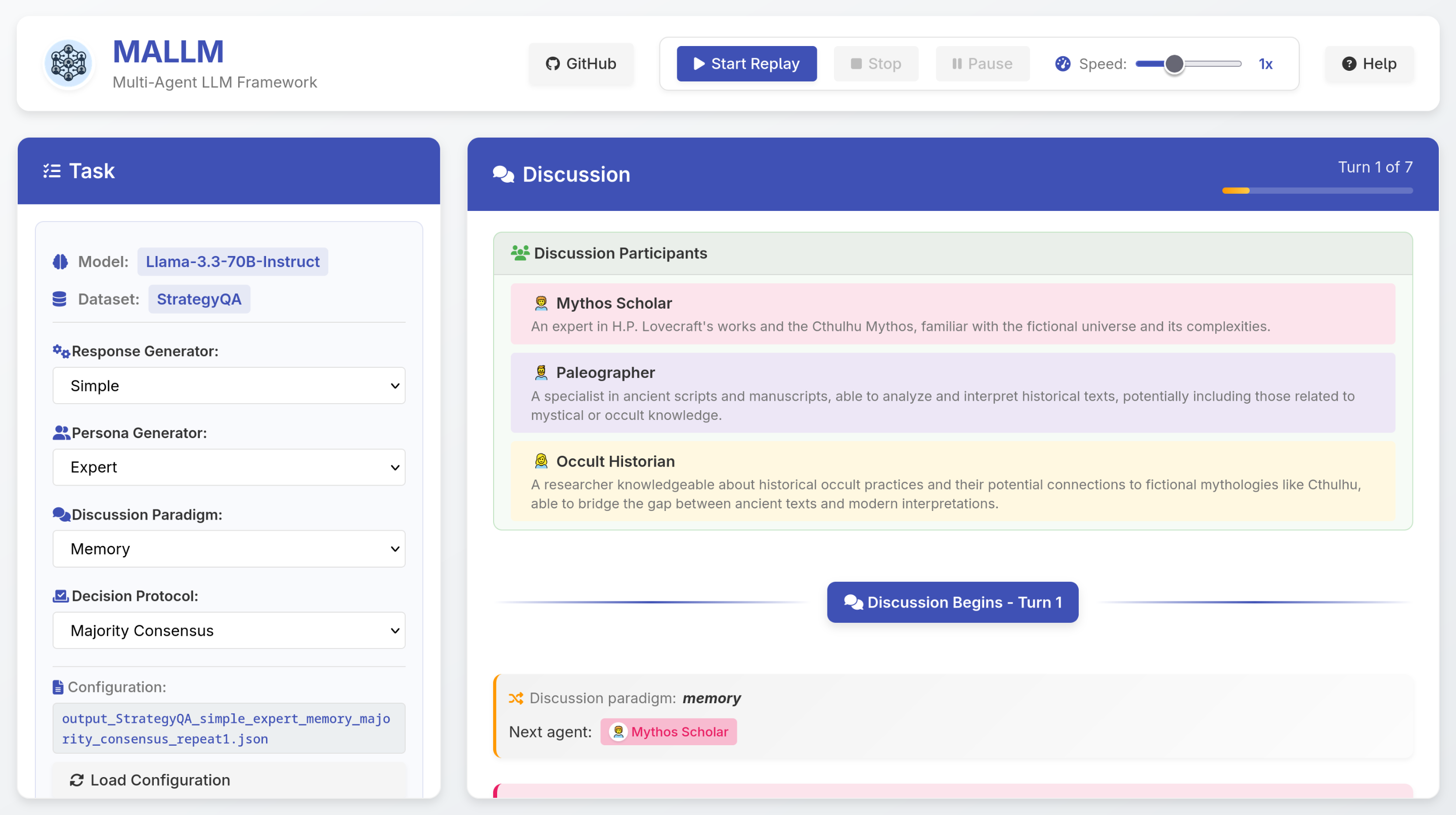}
    \caption{A screenshot of the demonstration website. One of 144 configurations for MAD can be selected on the left panel. MAD is conducted to solve the task, visible on the right panel. The top header provides functions to pause the replay or adjust the simulation speed.}
    \label{fig:app_mallm_screenshot}
\end{figure*}

We provide an interactive demonstration website for MALLM. It is available under \href{https://mallm.gipplab.org/}{mallm.gipplab.org}. A screenshot can be seen in \Cref{fig:app_mallm_screenshot}.

\section{Comparison With Other Frameworks} \label{app:other_frameworks}

We compare the functionality of other commonly used frameworks for MAD with MALLM. The comparison can be seen in \Cref{tab:frameworks}.

\section{DEBATE Examples} \label{app:dataset}
To demonstrate MALLM's capabilities, we construct a set of examples called DEBATE (\ul{D}iverse \ul{E}xchanges \ul{B}etween \ul{A}utonomous \ul{T}alking \ul{E}ntities), comprising 14,400 strategic problem-solving debates based on the StrategyQA dataset \citep{GevaKSK21}, generated using 144 distinct MALLM configurations. 
Each configuration combines specific settings of the framework's modular components listed in \Cref{tab:parameters}. 

\begin{table}[H]
\small
\centering
\begin{tabular}{l|p{4cm}} %
\toprule
\textbf{Parameter} & \textbf{Values} \\
\midrule
Response Generators & Simple, Critical, Reasoning \\
Persona Generators & None, Expert, IPIP \\
Discussion Paradigms & Memory, Relay, Report, Debate \\
Decision Protocols & Majority Consensus, Unanimity Consensus, Simple Voting, Approval Voting \\
\bottomrule
\end{tabular}
\caption{Parameters used for creating DEBATE. The example set comprises diverse data for each possible combination of parameters.}
\label{tab:parameters}
\end{table}

For example, one setup uses the \texttt{Simple} response generator, no personas, the \texttt{Memory} discussion paradigm, and the \texttt{Majority Consensus} decision protocol. 
Our rationale for selecting the parameters is to ensure diversity in the agent orchestration (e.g., two voting and two consensus approaches for decision-making), while keeping the computational effort manageable.
All debates involve three agents, up to seven turns, and use the \href{https://huggingface.co/meta-llama/Llama-3.3-70B-Instruct}{meta-llama/Llama-3.3-70B-Instruct} model.
Each of the 144 configurations runs 100 debates.
For the creation of the DEBATE examples, we utilize eight NVIDIA A100 GPUs, each equipped with 40GB of VRAM, to host a \href{https://huggingface.co/meta-llama/Llama-3.3-70B-Instruct}{meta-llama/Llama-3.3-70B-Instruct} model for 8 days, 5 hours, and 42 minutes.
The data is available on Hugging Face\footnote{\href{https://huggingface.co/datasets/Multi-Agent-LLMs/DEBATE/}{huggingface.co/datasets/Multi-Agent-LLMs/DEBATE}}.

\section{Parameters}\label{app:parameters}

We adhere to default parameters for the models we used, using langchain 0.1.16 and openai 1.25.0 for the implementation of the MALLM framework.

\begin{itemize}[noitemsep]
    \item \texttt{temperature = 1.0}
    \item \texttt{top\_p = 1.0}
    \item \texttt{presence\_penalty = 0.0}
    \item \texttt{frequency\_penalty = 0.0}
    \item \texttt{max\_tokens = 1024}
\end{itemize}

\begin{table*}[t]
\small
\centering
\begin{tabular}{lccccc}
\hline
\textbf{Customizable Feature} & \makecell{\textbf{Agent} \\ \textbf{Personas}} & \makecell{\textbf{Agent} \\ \textbf{Responses}} & \makecell{\textbf{Discussion} \\ \textbf{Paradigms}} & \makecell{\textbf{Decision} \\ \textbf{Protocol}} & \makecell{\textbf{Evaluation} \\ \textbf{Pipeline}} \\
\hline
AutoGen \citep{wu_autogen_2023} & \xmark & \xmark & \xmark & \xmark & \xmark \\
GPTSwarm \citep{zhuge_language_2024} & \xmark & \xmark & \cmark & \cmark & \xmark \\
OpenAI Agents SDK \citep{openai_agents_python} & \cmark & \xmark & \cmark & \xmark & \cmark \\
MetaGPT \citep{hong_metagpt_2024} & \cmark & \cmark & \xmark & \xmark & \xmark \\
AgentScope \citep{GaoLPK24} & \cmark & \xmark & \cmark & \xmark & \xmark \\
AutoGPT \citep{talebirad2023multiagentcollaborationharnessingpower} & \cmark & \cmark & \cmark & \xmark & \xmark \\
\hline
\textbf{MALLM (this work)} & \cmark & \cmark & \cmark & \cmark & \cmark \\
\hline
\end{tabular}
\caption{Comparison of customizable features across commonly used frameworks for MAD. MALLM enables the modification of agent personas, agent responses, discussion paradigms, and decision protocols. It also comes with an integrated evaluation pipeline. To the best of our knowledge, no other framework offers the same level of configurability for these main components of MAD.}
\label{tab:frameworks}
\end{table*}

\subsection{Agent Experiments}\label{app:agent_exp}

The setup and parameters for this experiment are described in \citet{kaesberg2025votingconsensusdecisionmakingmultiagent}.
They use \textit{meta-llama/Meta-Llama-3-8B-Instruct} as a model for all agents with the following fixed parameters:

\begin{itemize}
    \item Persona generator: \texttt{Expert}
    \item Discussion paradigm: \texttt{Memory}
    \item Decision protocol: Average of \texttt{Simple Voting}, \texttt{Ranked Voting}, \texttt{Cumulative Voting} and \texttt{Approval Voting}
\end{itemize}
Each experiment is repeated three times, and the average performance and standard deviation across the runs are reported.

\subsection{Discussion Experiments}\label{app:discussion_exp}

We use \textit{meta-llama/Meta-Llama-3-70B-Instruct} as a model for all agents.
We further report the parameters that are set fixed for this experiment:
\begin{itemize}
    \item Persona generator: \texttt{Expert}
    \item Response generator: \texttt{Simple}
    \item Decision protocol: \texttt{Majority Consensus}
\end{itemize}
To ensure the reliability of our findings, we follow the prior work of \citet{WangPSB24} and conduct each experiment five times, reporting the average performance and standard deviation across the runs.

\subsection{Decision Experiments}\label{app:decision_exp}

The setup and parameters for this experiment are described in \citet{kaesberg2025votingconsensusdecisionmakingmultiagent}.
They use \textit{meta-llama/Meta-Llama-3-8B-Instruct} as a model for all agents with the following fixed parameters:
\begin{itemize}
    \item Persona generator: \texttt{Expert}
    \item Discussion paradigm: \texttt{Memory}
    \item Response Generator: \texttt{Simple}
\end{itemize}
Each experiment is repeated three times, and the average performance and standard deviation across the runs are reported.

\section{Evaluation Pipeline}
\label{app:auto-charts}
\begin{figure}[H]
    \centering
    \includegraphics[width=0.95\linewidth]{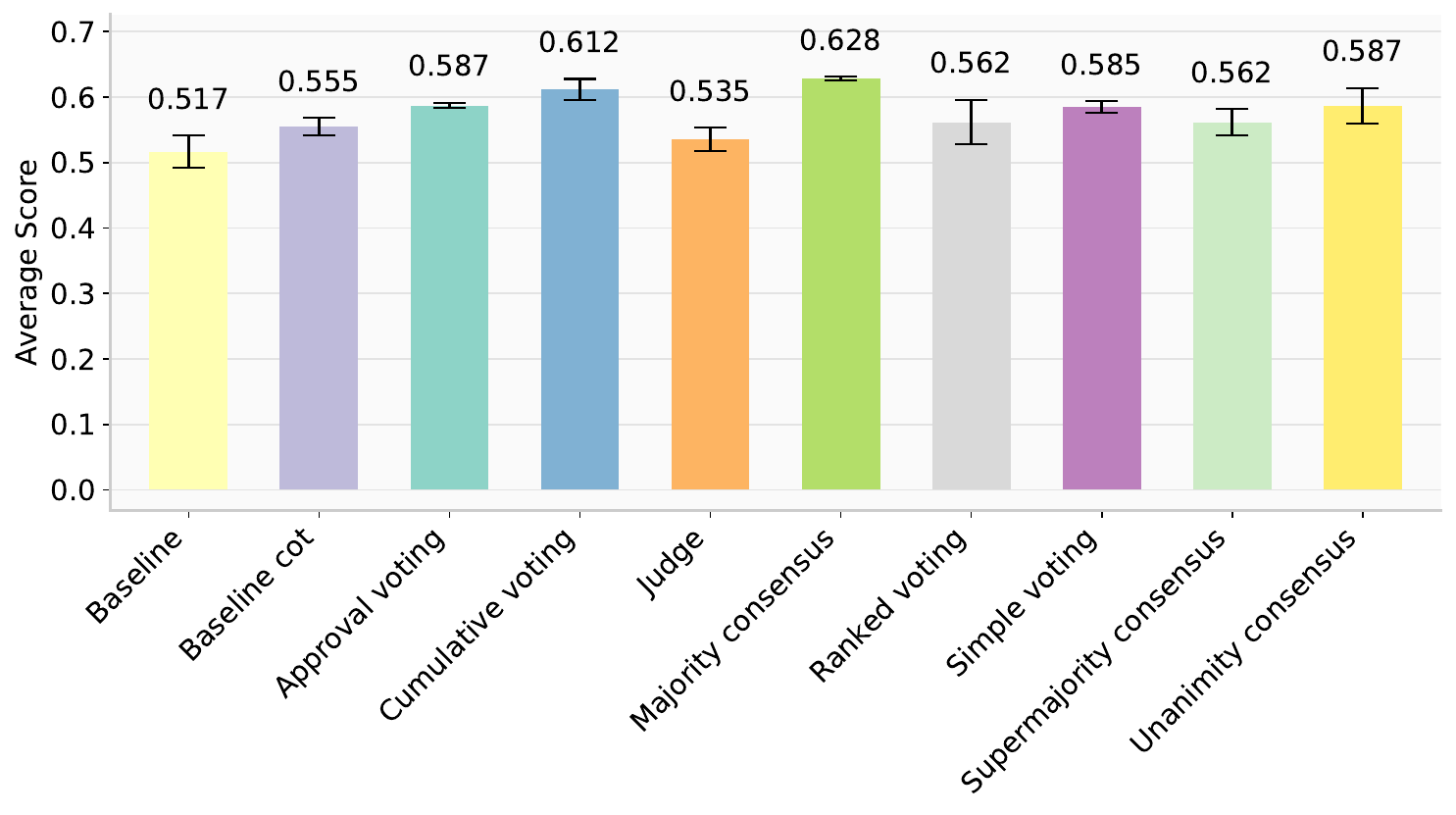}
    \caption{An example chart automatically generated by the MALLM evaluation pipeline, comparing the average performance scores of various decision protocols on the StrategyQA dataset. Error bars indicate the standard deviation over three experimental runs.}
    \label{fig:mallm-auto-score}
\end{figure}

\begin{figure}[H]
    \centering
    \includegraphics[width=0.95\linewidth]{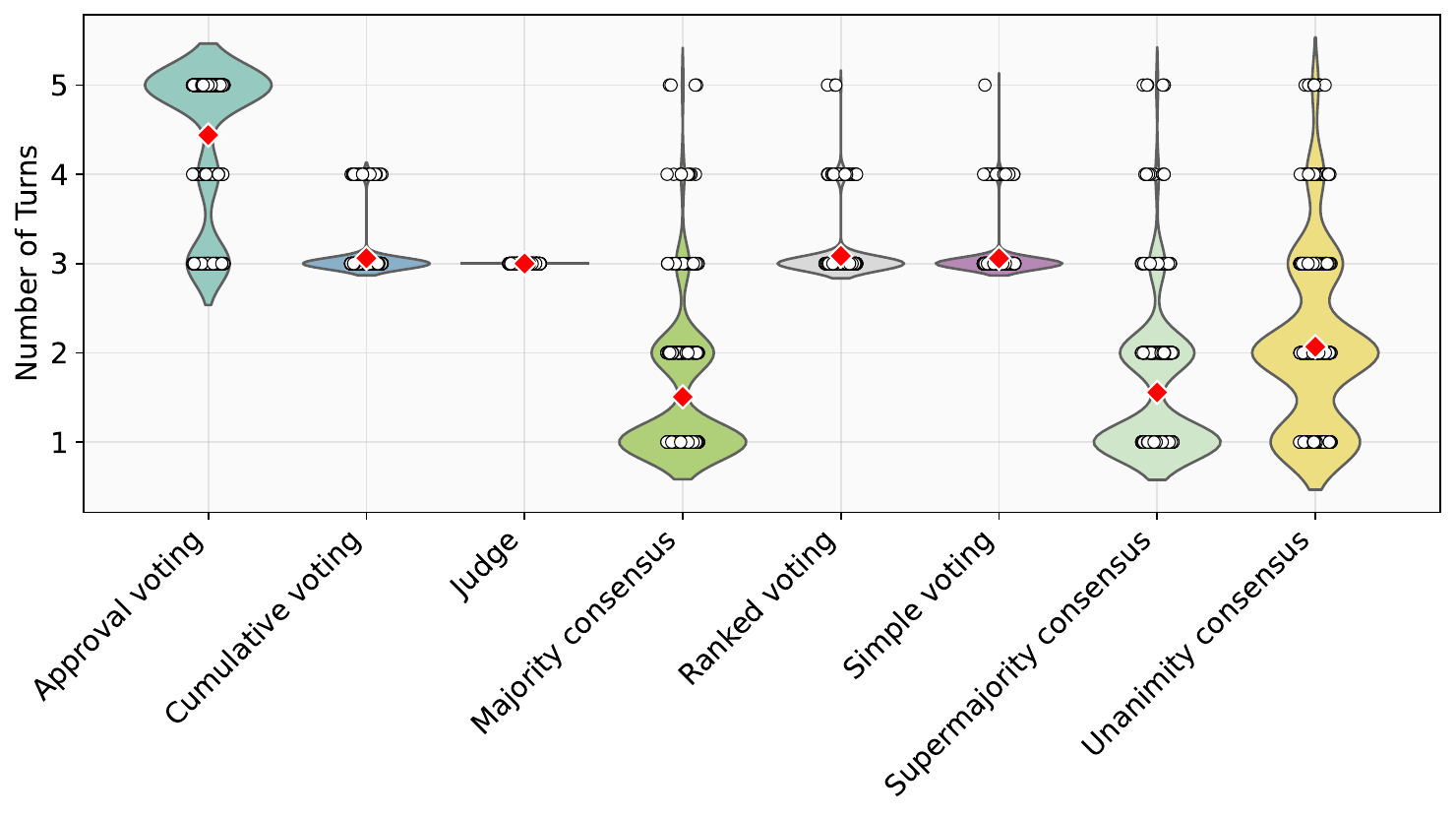}
    \caption{An example visualization from the MALLM evaluation pipeline, showing the distribution of the number of turns required for different decision protocols to converge on the StrategyQA dataset. The plot's width illustrates the frequency of turn counts, and the red marker shows the mean.}
    \label{fig:mallm-auto-turns}
\end{figure}

\begin{figure}[H]
    \centering
    \includegraphics[width=\linewidth]{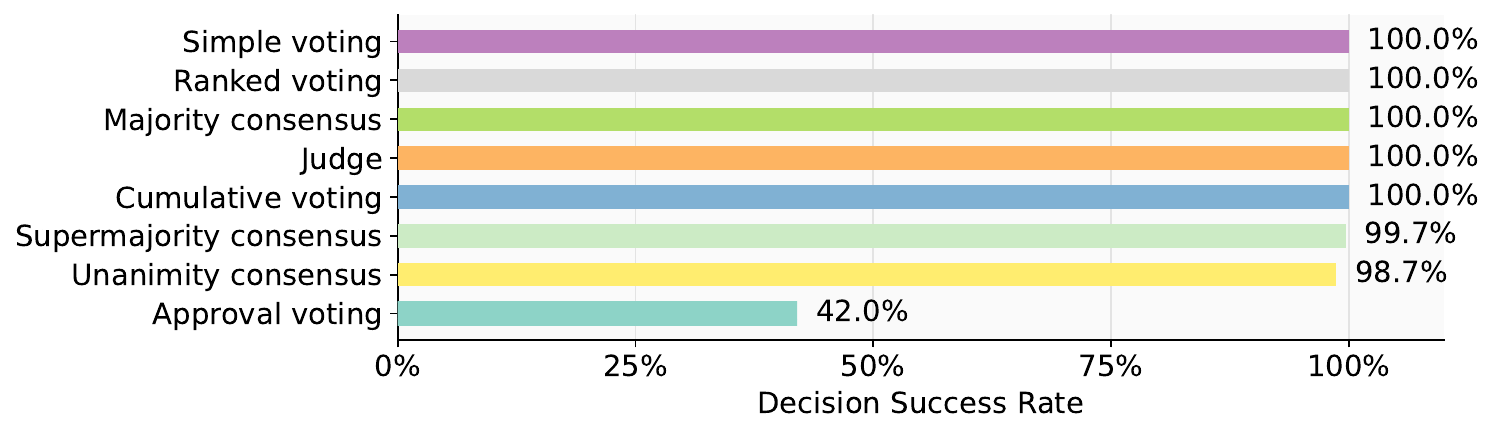}
    \caption{An example chart from the MALLM evaluation pipeline showing the decision success rates for each protocol on the StrategyQA dataset. The decision success rate explains how many of the debates reach a final solution according to the decision protocol (e.g., $>50\%$ for \texttt{Majority Consensus}).}
    \label{fig:mallm-auto-decision}
\end{figure}

\begin{figure}[H]
    \centering
    \includegraphics[width=\linewidth]{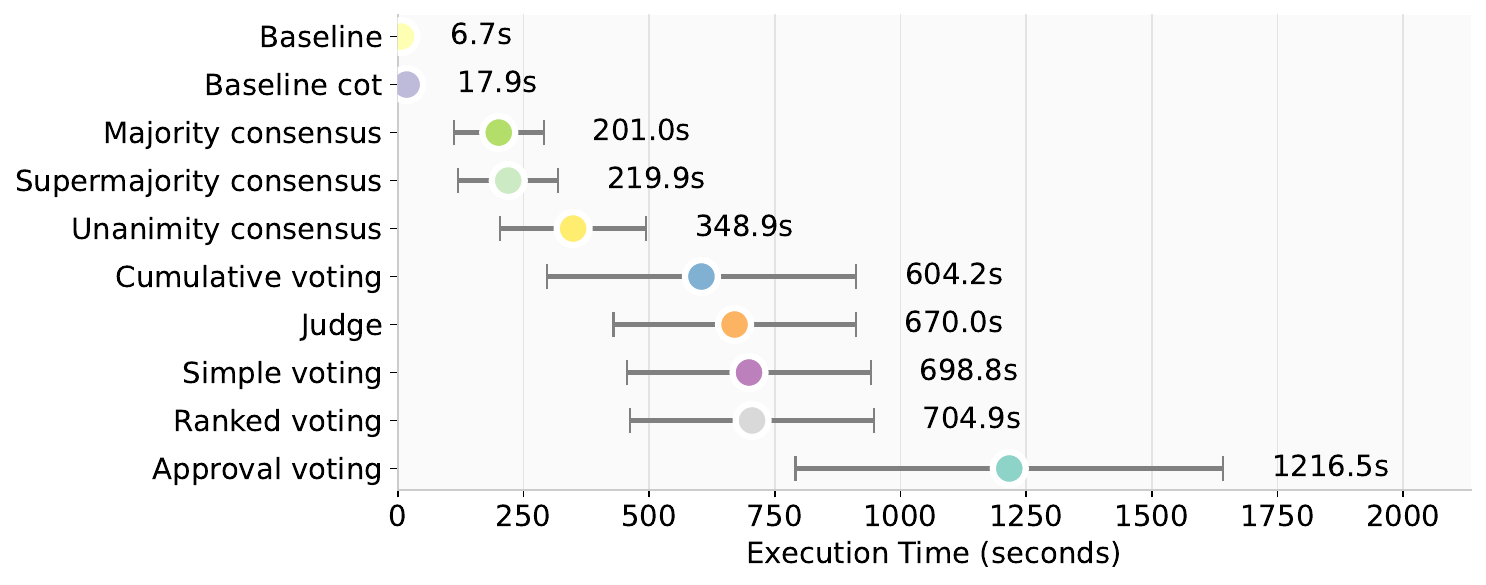}
    \caption{An example of an automatically generated chart from the MALLM evaluation pipeline, comparing the average wall clock time (in seconds) for each decision protocol on the StrategyQA dataset.}
    \label{fig:mallm-auto-clock}
\end{figure}

\Cref{fig:mallm-auto-clock,fig:mallm-auto-score,fig:mallm-auto-turns,fig:mallm-auto-decision} show example charts generated by the MALLM evaluation pipeline. They are presented to demonstrate the pipeline's automated analysis and visualization capabilities. These specific examples are generated from experiments on the StrategyQA dataset using the \textit{meta-llama/Meta-Llama-3-8B-Instruct} model, with error bars representing the standard deviation across three runs.
\onecolumn

\twocolumn
\section{Experiment Prompts}\label{app:prompts}

We provide the prompts that the MALLM framework uses to conduct MAD, which are relevant for our experiments. 
\Cref{sec:prompts_general} shows the prompt used across all configurations. Prompts specific to the persona experiments are provided in \Cref{sec:prompts_agent}, while prompts for the response generators are detailed in \Cref{app:prompts_responsegenerator}. Additionally, prompts related to the discussion paradigm are included in \Cref{app:discussion_prompt}, and prompts for the various decision protocols are available in \Cref{app:decision_prompt}.
    
\subsection{General Debate}\label{sec:prompts_general}

\begin{figure}[H]
    \centering
    \begin{combinedprompt}
    \textbf{System Prompt:} \\
    \begingroup
    \colorbox{systemcolor}{\parbox{\dimexpr\linewidth-2\fboxsep\relax}{
    You take part in a discussion to solve a task.\\
    \\
    Your role: \texttt{<persona name>} (\texttt{<persona description>})\\
    Task: \texttt{<instruction>}\\
    Input: \texttt{<example>}\\
    Context: \texttt{<optional information>}\\
    Current Solution: \texttt{<most recent draft>}\\
    Discussion so far: \texttt{<agent memory>}
    }}
    \endgroup
\end{combinedprompt}
    \caption{Prompt used with the \texttt{Simple} response generator for an agent participating in collaborative debate. If this is the first message of the discussion, we write ``Nobody proposed a solution yet. Please provide the first one.'' instead of the most recent draft and agent memory.}
    \label{fig:discussion_prompt}
\end{figure}

\subsection{Persona Experiments} 
\label{sec:prompts_agent}

These are the prompts used for the persona experiments.

\begin{figure}[H]
    \centering
    \begin{combinedprompt}
    \textbf{System Prompt:} \\
    \begingroup
    \colorbox{systemcolor}{\parbox{\dimexpr\linewidth-2\fboxsep\relax}{
    Solve the following task: \texttt{<task instruction>} \\ Input: \texttt{<input str>} \\ Make absolutely sure to provide your solution in the end: 'FINAL SOLUTION: <Letter>'.
    }}
    \endgroup

    \vspace{0.4em} %

    \textbf{User Prompt:} \\
    \begingroup
    \colorbox{usercolor}{\parbox{\dimexpr\linewidth-2\fboxsep\relax}{
    Answer the following question.
    }}
    \endgroup
\end{combinedprompt}
    \caption{Base prompt used for agents participating in a GPQA experiment.}
\end{figure}

\begin{figure}[H]
    \centering
    \begin{combinedprompt}
    \textbf{System Prompt:} \\
    \begingroup
    \colorbox{systemcolor}{\parbox{\dimexpr\linewidth-2\fboxsep\relax}{
    When faced with a task, begin by identifying the participants who will contribute to solving the task. Provide role and description of the participants, describing their expertise or needs, formatted using the provided JSON schema. \\
    Generate one participant at a time, complementing the existing participants to foster a rich discussion.

    Example 1: \texttt{<example 1>}
    
    Example 2: \texttt{<example 2>}
    
    Example 3: \texttt{<example 3>}
    }}
    \endgroup

    \vspace{0.4em} %

    \textbf{User Prompt:} \\
    \begingroup
    \colorbox{usercolor}{\parbox{\dimexpr\linewidth-2\fboxsep\relax}{
    Now generate a participant to discuss the following task:\\Task: \texttt{<task description>}.
    Please use the following examples to generate a useful persona for the task! Only answer with the JSON for the next persona.}}
    \endgroup
\end{combinedprompt}
    \caption{Prompt used for the Expert agent generator, which creates unique personas for each example.}
\end{figure}

\begin{figure}[H]
    \centering
    \begin{combinedprompt}
    \textbf{System Prompt:} \\
    \begingroup
    \colorbox{systemcolor}{\parbox{\dimexpr\linewidth-2\fboxsep\relax}{
    When faced with a task, begin by identifying the participants who will contribute to solving the task. Provide role and fixed characteristics of the participant, formatted using the provided JSON schema.
    Generate one participant at a time, complementing the existing participants to foster a rich discussion.
    
    You must choose the following characteristics for the participant, in JSON format:
    \\ \texttt{<characteristics and options>}
    
    You absolutely must stick to the JSON format and the characteristics and options provided.
    
    Example 1: \texttt{<example 1>}
    
    Example 2: \texttt{<example 2>}
    }}
    \endgroup

    \vspace{0.4em} %

    \textbf{User Prompt:} \\
    \begingroup
    \colorbox{usercolor}{\parbox{\dimexpr\linewidth-2\fboxsep\relax}{
    Now generate a participant to discuss the following task:\\Task: \texttt{<task description>}.
    Only answer with the JSON for the next persona! Ensure your new participant is unique.}}
    \endgroup
\end{combinedprompt}
    \caption{Prompt used for the IPIP agent generator, which creates unique personas for each example.}
\end{figure}

\subsection{Response Generator Experiments} \label{app:prompts_responsegenerator}
We provide the prompts for each response generator used.

\subsubsection{Simple Response Generator}
\begin{figure}[H]
    \centering
    \begin{combinedprompt}
    \textbf{User Prompt:} \\
    \begingroup
    \colorbox{usercolor}{\parbox{\dimexpr\linewidth-2\fboxsep\relax}{
    Based on the provided feedback, carefully re-examine your previous solution. Provide a revised solution.
    }}
    \endgroup
\end{combinedprompt}
    \caption{Prompt with the \texttt{Simple} response generator instructing an agent to create a new solution based on received feedback. It is appended to the system prompt in \Cref{fig:discussion_prompt}.}
    \label{fig:user_revision_prompt}
\end{figure}

\begin{figure}[H]
    \centering
    \begin{combinedprompt}
    \textbf{User Prompt:} \\
    \begingroup
    \colorbox{usercolor}{\parbox{\dimexpr\linewidth-2\fboxsep\relax}{
    Improve the current solution.\\
    If you agree with the current solution, answer with \texttt{[AGREE]}.\\
    Else, answer with \texttt{[DISAGREE]}, explain why, and provide an improved solution.\\
    Let's think step-by-step.
    }}
    \endgroup
\end{combinedprompt}
    \caption{Prompt with the \texttt{Simple} response generator to an agent for contributing to the current solution draft. The agent can either agree or disagree. It is appended to the system prompt in \Cref{fig:discussion_prompt}.}
    \label{fig:user_only_improvement_prompt}
\end{figure}

\begin{figure}[H]
    \centering
    \begin{combinedprompt}
    \textbf{User Prompt:} \\
    \begingroup
    \colorbox{usercolor}{\parbox{\dimexpr\linewidth-2\fboxsep\relax}{
    Improve the current solution.\\
    Based on the current solution, give constructive feedback. If you agree, answer with \texttt{[AGREE]}, else answer with \texttt{[DISAGREE]} and explain why.\\
    Let's think step-by-step.
    }}
    \endgroup
\end{combinedprompt}
    \caption{Prompt with the \texttt{Simple} response generator to an agent for giving feedback to the current solution draft (without directly proposing a new solution). The agent can either agree or disagree. It is appended to the system prompt in \Cref{fig:discussion_prompt}.}
    \label{fig:user_only_feedback_prompt}
\end{figure}

\subsubsection{Critical Response Generator}
\begin{figure}[H]
    \centering
    \begin{combinedprompt}
    \textbf{User Prompt:} \\
    \begingroup
    \colorbox{usercolor}{\parbox{\dimexpr\linewidth-2\fboxsep\relax}{
    Re-examine the current solution critically based on the feedback provided. Ensure your revision addresses any identified weaknesses or areas for improvement. Submit a revised and improved solution.
    }}
    \endgroup
\end{combinedprompt}
    \caption{Prompt with the \texttt{Critical} response generator instructing an agent to create a new solution based on received feedback. It is appended to the system prompt in \Cref{fig:discussion_prompt}.}
    \label{fig:user_revision_prompt_critical}
\end{figure}

\begin{figure}[H]
    \centering
    \begin{combinedprompt}
    \textbf{User Prompt:} \\
    \begingroup
    \colorbox{usercolor}{\parbox{\dimexpr\linewidth-2\fboxsep\relax}{
    Improve the current solution. Identify specific areas that need enhancement and propose unique solutions based on your persona. If you see no room for improvement, answer with \texttt{[AGREE]}, otherwise, answer with \texttt{[DISAGREE]} and provide a clear, solution.
    }}
    \endgroup
\end{combinedprompt}
    \caption{Prompt with the \texttt{Critical} response generator to an agent for contributing to the current solution draft. The agent can either agree or disagree. It is appended to the system prompt in \Cref{fig:discussion_prompt}.}
    \label{fig:user_only_improvement_prompt_critical}
\end{figure}

\begin{figure}[H]
    \centering
    \begin{combinedprompt}
    \textbf{User Prompt:} \\
    \begingroup
    \colorbox{usercolor}{\parbox{\dimexpr\linewidth-2\fboxsep\relax}{
    Critically evaluate the current solution. Identify potential weaknesses or areas of improvement. If you believe the solution is flawless, answer with \texttt{[AGREE]}, otherwise answer with \texttt{[DISAGREE]} and provide constructive feedback with suggestions for improvement.
    }}
    \endgroup
\end{combinedprompt}
    \caption{Prompt with the \texttt{Critical} response generator to an agent for giving feedback to the current solution draft (without directly proposing a new solution). The agent can either agree or disagree. It is appended to the system prompt in \Cref{fig:discussion_prompt}.}
    \label{fig:user_only_feedback_prompt_critical}
\end{figure}

\subsubsection{Reasoning Response Generator}
\begin{figure}[H]
    \centering
    \begin{combinedprompt}
    \textbf{User Prompt:} \\
    \begingroup
    \colorbox{usercolor}{\parbox{\dimexpr\linewidth-2\fboxsep\relax}{
    Based on the provided feedback, carefully re-examine your previous solution. Provide a revised solution.
    }}
    \endgroup
\end{combinedprompt}
    \caption{Prompt with the \texttt{Reasoning} response generator instructing an agent to create a new solution based on received feedback. It is appended to the system prompt in \Cref{fig:discussion_prompt}.}
    \label{fig:user_revision_prompt_reasoning}
\end{figure}

\begin{figure}[H]
    \centering
    \begin{combinedprompt}
    \textbf{User Prompt:} \\
    \begingroup
    \colorbox{usercolor}{\parbox{\dimexpr\linewidth-2\fboxsep\relax}{
    Improve the current steps of the argument by referring to the other participants in the discussion. Be critical and answer short and concise. Repeat only the reasoning steps that you think are the most important. If you think there is enough information to create a final answer also answer with \texttt{[AGREE]} else answer with \texttt{[DISAGREE]}. Don't provide a final solution yet.
    }}
    \endgroup
\end{combinedprompt}
    \caption{Prompt with the \texttt{Reasoning} response generator to an agent for contributing to the current solution draft. The agent can either agree or disagree. It is appended to the system prompt in \Cref{fig:discussion_prompt}.}
    \label{fig:user_only_improvement_prompt_reasoning}
\end{figure}

\begin{figure}[H]
    \centering
    \begin{combinedprompt}
    \textbf{User Prompt:} \\
    \begingroup
    \colorbox{usercolor}{\parbox{\dimexpr\linewidth-2\fboxsep\relax}{
    Based on the current solution, give constructive feedback. If you agree, answer with \texttt{[AGREE]}, else answer with \texttt{[DISAGREE]} and explain why.
    }}
    \endgroup
\end{combinedprompt}
    \caption{Prompt with the \texttt{Reasoning} response generator to an agent for giving feedback to the current solution draft (without directly proposing a new solution). The agent can either agree or disagree. It is appended to the system prompt in \Cref{fig:discussion_prompt}.}
    \label{fig:user_only_feedback_prompt_reasoning}
\end{figure}

\subsection{Discussion Paradigm Experiments}
\label{app:discussion_prompt}
For the experiments on discussion paradigms, just one more prompt for the Chain-of-Thought baseline is used. We use the general prompts described in \Cref{sec:prompts_general} for MAD.

\begin{figure}[H]
    \centering
    \begin{combinedprompt}
    \textbf{System Prompt:} \\
    \begingroup
    \colorbox{systemcolor}{\parbox{\dimexpr\linewidth-2\fboxsep\relax}{
    Solve the provided task. Do not ask back questions. Clearly indicate your final solution after the text 'Final Solution:'.\\
    \\
    Task: \texttt{<task instruction>} \\
    Input: \texttt{<input str>}
    }}
    \endgroup

    \vspace{0.4em} %

    \textbf{User Prompt:} \\
    \begingroup
    \colorbox{usercolor}{\parbox{\dimexpr\linewidth-2\fboxsep\relax}{
    Let's think step-by-step.
    }}
    \endgroup
\end{combinedprompt}
    \caption{Prompt used for the Chain-of-Thought baseline.}
    \label{fig:cot_prompt}
\end{figure}

\subsection{Decision Protocol Experiments}
\label{app:decision_prompt}
These are all prompts used for the decision-making protocols. The final answer extraction prompt can be seen in \Cref{fig:extract_prompt}. The prompt for the voting-based decision protocols can be seen in \Cref{fig:simple_voting_prompt} (Simple Voting), \Cref{fig:approval_voting_prompt} (Approval Voting), \Cref{fig:ranked_voting_prompt} (Ranked Voting), and \Cref{fig:cumulative_voting_prompt} (Cumulative Voting) 
decision protocols (\Cref{fig:simple_voting_prompt} to \Cref{fig:judge_prompt}). The consensus decision protocol has no special prompt, as it terminates when a consensus is found, and then the final answer extraction prompt is used. Voting also utilizes the final answer extraction prompt to obtain the final answer from each agent that is used during the voting process.

\subsubsection{Final Answer Extraction}
\label{sec:app_final_answer_prompt}
\medskip
\begin{figure}[H]
    \centering
    \begin{combinedprompt}
    \textbf{System Prompt:} \\
    \begingroup
    \colorbox{systemcolor}{\parbox{\dimexpr\linewidth-2\fboxsep\relax}{
    Your role: \texttt{<persona>} (\texttt{<persona description>})
    }}
    \endgroup

    \vspace{0.4em} %

    \textbf{User Prompt:} \\
    \begingroup
    \colorbox{usercolor}{\parbox{\dimexpr\linewidth-2\fboxsep\relax}{
    You are tasked with creating a final solution based on the given input and your previous response.\\
    Task: \texttt{<task>}\\
    Input: \texttt{<input sample>}\\
    Your previous response: \texttt{<previous answer>}\\
    Extract the final solution to the task from the provided text. Remove statements of agreement, disagreement, and explanations. Do not modify the text. Do not output any text besides the solution. If there is no solution provided, just copy the previous response.
    }}
    \endgroup
\end{combinedprompt}
    \caption{Prompt used to extract the final answer of a given agent from its previous response.}
    \label{fig:extract_prompt}
\end{figure}

\subsubsection{Voting Prompts}
\label{app:voting_prompts}

\begin{figure}[H]
    \centering
    \begin{combinedprompt}
    \textbf{System Prompt:} \\
    \begingroup
    \colorbox{systemcolor}{\parbox{\dimexpr\linewidth-2\fboxsep\relax}{
    Your role: \texttt{<persona>} (\texttt{<persona description>})
    }}
    \endgroup

    \vspace{0.4em} %

    \textbf{User Prompt:} \\
    \begingroup
    \colorbox{usercolor}{\parbox{\dimexpr\linewidth-2\fboxsep\relax}{
    You are tasked with voting for the best solution from the list provided below based on the given task.\\
    Task: \texttt{<task>}\\
    Question: \texttt{<input sample>}\\
    Here are the possible solutions:\\
    Solution 1: \texttt{<agent 1 final answer>}\\
    Solution 2: \texttt{<agent 2 final answer>}\\
    Solution 3: \texttt{<agent 3 final answer>}\\
    Based on the above solutions, please provide the number of the solution you are voting for. Answer only with the number.
    }}
    \endgroup
\end{combinedprompt}
    \caption{Prompt used to get a vote from each agent for the Simple Voting decision protocol.}
    \label{fig:simple_voting_prompt}
\end{figure}

\begin{figure}[H]
    \centering
    \begin{combinedprompt}
    \textbf{System Prompt:} \\
    \begingroup
    \colorbox{systemcolor}{\parbox{\dimexpr\linewidth-2\fboxsep\relax}{
    Your role: \texttt{<persona>} (\texttt{<persona description>})
    }}
    \endgroup

    \vspace{0.4em} %

    \textbf{User Prompt:} \\
    \begingroup
    \colorbox{usercolor}{\parbox{\dimexpr\linewidth-2\fboxsep\relax}{
    You are tasked with approving any number of solutions from the list provided below based on the given task.\\
    Task: \texttt{<task>}\\
    Question: \texttt{<input sample>}\\
    Here are the possible solutions:\\
    Solution 1: \texttt{<agent 1 final answer>}\\
    Solution 2: \texttt{<agent 2 final answer>}\\
    Solution 3: \texttt{<agent 3 final answer>}\\
    Based on the above solutions, please provide the numbers of the solutions you are approving, separated by commas. Answer only with the numbers.
    }}
    \endgroup
\end{combinedprompt}
    \caption{Prompt used to get a vote from each agent for the Approval Voting decision protocol.}
    \label{fig:approval_voting_prompt}
\end{figure}

\begin{figure}[H]
    \centering
    \begin{combinedprompt}
    \textbf{System Prompt:} \\
    \begingroup
    \colorbox{systemcolor}{\parbox{\dimexpr\linewidth-2\fboxsep\relax}{
    Your role: \texttt{<persona>} (\texttt{<persona description>})
    }}
    \endgroup

    \vspace{0.4em} %

    \textbf{User Prompt:} \\
    \begingroup
    \colorbox{usercolor}{\parbox{\dimexpr\linewidth-2\fboxsep\relax}{
    You are tasked with distributing 10 points among the provided solutions based on the given task.\\
    Task: \texttt{<task>}\\
    Question: \texttt{<input sample>}\\
    Here are the possible solutions:\\
    Solution 1: \texttt{<agent 1 final answer>}\\
    Solution 2: \texttt{<agent 2 final answer>}\\
    Solution 3: \texttt{<agent 3 final answer>}\\
    Based on the above solutions, please distribute 10 points among the solutions. Provide your points allocation as a JSON dictionary where keys are solution numbers (as int) and values are the points. The total points should sum up to 10. Answer only with the JSON dictionary.
    }}
    \endgroup
\end{combinedprompt}
    \caption{Prompt used to get a vote from each agent for the Cumulative Voting decision protocol.}
    \label{fig:cumulative_voting_prompt}
\end{figure}

\begin{figure}[H]
    \centering
    \begin{combinedprompt}
    \textbf{System Prompt:} \\
    \begingroup
    \colorbox{systemcolor}{\parbox{\dimexpr\linewidth-2\fboxsep\relax}{
    Your role: \texttt{<persona>} (\texttt{<persona description>})
    }}
    \endgroup

    \vspace{0.4em} %

    \textbf{User Prompt:} \\
    \begingroup
    \colorbox{usercolor}{\parbox{\dimexpr\linewidth-2\fboxsep\relax}{
    You are tasked with ranking the solutions from the most preferred to the least preferred based on the given task.\\
    Task: \texttt{<task>}\\
    Question: \texttt{<input sample>}\\
    Here are the possible solutions:\\
    Solution 1: \texttt{<agent 1 final answer>}\\
    Solution 2: \texttt{<agent 2 final answer>}\\
    Solution 3: \texttt{<agent 3 final answer>}\\
    Based on the above solutions, please provide the rankings of the solutions separated by spaces. Example: '0 2 1' if you prefer Solution 0 the most, then Solution 2, and finally Solution 1. Provide up to 5 rankings. Only answer with the rankings.
    }}
    \endgroup
\end{combinedprompt}
    \caption{Prompt used to get a vote from each agent for the Ranked Voting decision protocol.}
    \label{fig:ranked_voting_prompt}
\end{figure}

\subsubsection{Judge Prompt}

\begin{figure}[H]
    \centering
    \begin{combinedprompt}
    \textbf{User Prompt:} \\
    \begingroup
    \colorbox{usercolor}{\parbox{\dimexpr\linewidth-2\fboxsep\relax}{
    Task: \texttt{<task>}\\
    Question: \texttt{<input sample>}\\
    Please provide a decision on the following solutions and combine them in a single answer to solve the task. Only answer with the solution:\\
    Solution 1: \texttt{<agent 1 final answer>}\\
    Solution 2: \texttt{<agent 2 final answer>}\\
    Solution 3: \texttt{<agent 3 final answer>}
    }}
    \endgroup
\end{combinedprompt}
    \caption{Prompt used to get a final decision from the Judge decision protocol. No alterations are applied.}
    \label{fig:judge_prompt}
\end{figure}

\section{MALLM Configuration File}
\label{app:batch-config-file}

Example MALLM batch configuration file for running an experiment with fixed response generator, persona generator, and discussion paradigm, but varying decision protocols. The "repeats" field defines how many times each run is repeated, which is later relevant for evaluating for robustness by the standard deviation between experiment runs. The "common" field describes parameters that are considered for all runs. The "runs" field defines the parameters unique for each individual run.
\onecolumn
  \begin{minipage}{\linewidth}
    \begin{lstlisting}
{
  "repeats": 3,
  "name": "<DATASET NAME>",
  "common": {
    "task_instruction_prompt_template": "<DATASET NAME>",
    "endpoint_url": "<LLM API HOSTNAME>",
    "api_key": "<LLM API KEY>",
    "model_name": "<MODEL NAME>",
    "input_json_file_path": "data/datasets/<DATASET NAME>.json",
    "concurrent_api_requests": 200,
    "num_samples": "<NUMBER OF SAMPLES>",
    "max_turns": 5,
    "response_generator":"simple"
  },
  "runs": [
    {
      "output_json_file_path": "results/baseline-cot.json",
      "use_baseline": true
    },
    {
      "output_json_file_path": "results/baseline.json",
      "use_baseline": true,
      "use_chain_of_thought": false
    },
    {
      "output_json_file_path": "results/approval.json",
      "decision_protocol": "approval_voting"
    },
    {
      "output_json_file_path": "results/cumulative.json",
      "decision_protocol": "cumulative_voting"
    },
    {
      "output_json_file_path": "results/majority_consensus.json",
      "decision_protocol": "majority_consensus"
    },
    {
      "output_json_file_path": "results/supermajority_consensus.json",
      "decision_protocol": "supermajority_consensus"
    },
    {
      "output_json_file_path": "results/unanimity_consensus.json",
      "decision_protocol": "unanimity_consensus"
    },
    {
      "output_json_file_path": "results/voting.json",
      "decision_protocol": "simple_voting"
    },
    {
      "output_json_file_path": "results/ranked.json",
      "decision_protocol": "ranked_voting"
    }
  ]
}
    \end{lstlisting}
  \end{minipage}

\onecolumn

\clearpage
\onecolumn
\hypertarget{annotation}{}
\pagestyle{empty}
\lstset{
  basicstyle=\footnotesize\ttfamily,
  breaklines=true,
  breakatwhitespace=false,
  columns=flexible,
  numbers=none
}

\definecolor{Primary}{RGB}{59, 130, 246}    %
\definecolor{PrimaryDark}{RGB}{30, 64, 175} %
\definecolor{LightBg}{RGB}{239, 246, 255}   %
\definecolor{TextDark}{RGB}{31, 41, 55}     %
\definecolor{TextMuted}{RGB}{107, 114, 128} %

\begin{tikzpicture}[remember picture, overlay]
  \fill[Primary] ([xshift=0cm,yshift=0cm]current page.north west) rectangle ([xshift=\paperwidth,yshift=-0.4cm]current page.north west);
\end{tikzpicture}

\vspace{0.8cm}
\begin{center}
  {\fontsize{22}{26}\selectfont\sffamily\bfseries \textcolor{PrimaryDark}{CiteAssist}}\\[0.2em]
  {\Large\sffamily\scshape \textcolor{TextMuted}{Citation Sheet}}\\[0.8em]
  {\small\sffamily Generated with \href{https://citeassist.uni-goettingen.de/}{\textcolor{Primary}{\texttt{citeassist.uni-goettingen.de}}}
  }\end{center}

\begin{center}
\vspace{1em}
\begin{tikzpicture}
\draw[Primary, line width=0.6pt] (0,0) -- (\textwidth,0);
\end{tikzpicture}
\vspace{1.2em}
\end{center}

\begin{tcolorbox}[enhanced,
                 frame hidden,
                 boxrule=0pt,
                 borderline west={2pt}{0pt}{Primary},
                 colback=LightBg,
                 sharp corners,
                 breakable,
                 fonttitle=\sffamily\bfseries\large,
                 coltitle=Primary,
                 title=BibTeX Entry,
                 attach title to upper={\vspace{0.2em}\par},
                 left=12pt]
\lstset{
    inputencoding = utf8,  %
    extendedchars = true,  %
    literate      =        %
      {á}{{\'a}}1  {é}{{\'e}}1  {í}{{\'i}}1 {ó}{{\'o}}1  {ú}{{\'u}}1
      {Á}{{\'A}}1  {É}{{\'E}}1  {Í}{{\'I}}1 {Ó}{{\'O}}1  {Ú}{{\'U}}1
      {à}{{\`a}}1  {è}{{\`e}}1  {ì}{{\`i}}1 {ò}{{\`o}}1  {ù}{{\`u}}1
      {À}{{\`A}}1  {È}{{\`E}}1  {Ì}{{\`I}}1 {Ò}{{\`O}}1  {Ù}{{\`U}}1
      {ä}{{\"a}}1  {ë}{{\"e}}1  {ï}{{\"i}}1 {ö}{{\"o}}1  {ü}{{\"u}}1
      {Ä}{{\"A}}1  {Ë}{{\"E}}1  {Ï}{{\"I}}1 {Ö}{{\"O}}1  {Ü}{{\"U}}1
      {â}{{\^a}}1  {ê}{{\^e}}1  {î}{{\^i}}1 {ô}{{\^o}}1  {û}{{\^u}}1
      {Â}{{\^A}}1  {Ê}{{\^E}}1  {Î}{{\^I}}1 {Ô}{{\^O}}1  {Û}{{\^U}}1
      {œ}{{\oe}}1  {Œ}{{\OE}}1  {æ}{{\ae}}1 {Æ}{{\AE}}1  {ß}{{\ss}}1
      {ẞ}{{\SS}}1  {ç}{{\c{c}}}1 {Ç}{{\c{C}}}1 {ø}{{\o}}1  {Ø}{{\O}}1
      {å}{{\aa}}1  {Å}{{\AA}}1  {ã}{{\~a}}1  {õ}{{\~o}}1 {Ã}{{\~A}}1
      {Õ}{{\~O}}1  {ñ}{{\~n}}1  {Ñ}{{\~N}}1  {¿}{{?\`}}1  {¡}{{!\`}}1
      {„}{\quotedblbase}1 {“}{\textquotedblleft}1 {–}{$-$}1
      {°}{{\textdegree}}1 {º}{{\textordmasculine}}1 {ª}{{\textordfeminine}}1
      {£}{{\pounds}}1  {©}{{\copyright}}1  {®}{{\textregistered}}1
      {«}{{\guillemotleft}}1  {»}{{\guillemotright}}1  {Ð}{{\DH}}1  {ð}{{\dh}}1
      {Ý}{{\'Y}}1    {ý}{{\'y}}1    {Þ}{{\TH}}1    {þ}{{\th}}1    {Ă}{{\u{A}}}1
      {ă}{{\u{a}}}1  {Ą}{{\k{A}}}1  {ą}{{\k{a}}}1  {Ć}{{\'C}}1    {ć}{{\'c}}1
      {Č}{{\v{C}}}1  {č}{{\v{c}}}1  {Ď}{{\v{D}}}1  {ď}{{\v{d}}}1  {Đ}{{\DJ}}1
      {đ}{{\dj}}1    {Ė}{{\.{E}}}1  {ė}{{\.{e}}}1  {Ę}{{\k{E}}}1  {ę}{{\k{e}}}1
      {Ě}{{\v{E}}}1  {ě}{{\v{e}}}1  {Ğ}{{\u{G}}}1  {ğ}{{\u{g}}}1  {Ĩ}{{\~I}}1
      {ĩ}{{\~\i}}1   {Į}{{\k{I}}}1  {į}{{\k{i}}}1  {İ}{{\.{I}}}1  {ı}{{\i}}1
      {Ĺ}{{\'L}}1    {ĺ}{{\'l}}1    {Ľ}{{\v{L}}}1  {ľ}{{\v{l}}}1  {Ł}{{\L{}}}1
      {ł}{{\l{}}}1   {Ń}{{\'N}}1    {ń}{{\'n}}1    {Ň}{{\v{N}}}1  {ň}{{\v{n}}}1
      {Ő}{{\H{O}}}1  {ő}{{\H{o}}}1  {Ŕ}{{\'{R}}}1  {ŕ}{{\'{r}}}1  {Ř}{{\v{R}}}1
      {ř}{{\v{r}}}1  {Ś}{{\'S}}1    {ś}{{\'s}}1    {Ş}{{\c{S}}}1  {ş}{{\c{s}}}1
      {Š}{{\v{S}}}1  {š}{{\v{s}}}1  {Ť}{{\v{T}}}1  {ť}{{\v{t}}}1  {Ũ}{{\~U}}1
      {ũ}{{\~u}}1    {Ū}{{\={U}}}1  {ū}{{\={u}}}1  {Ů}{{\r{U}}}1  {ů}{{\r{u}}}1
      {Ű}{{\H{U}}}1  {ű}{{\H{u}}}1  {Ų}{{\k{U}}}1  {ų}{{\k{u}}}1  {Ź}{{\'Z}}1
      {ź}{{\'z}}1    {Ż}{{\.Z}}1    {ż}{{\.z}}1    {Ž}{{\v{Z}}}1  {ž}{{\v{z}}}1
  }
\begin{lstlisting}
@inproceedings{becker2025,
  author={Becker, Jonas and Kaesberg, Lars Benedikt and Bauer, Niklas and Wahle, Jan Philip and Ruas, Terry and Gipp, Bela},
  title={{MALLM}: Multi-Agent Large Language Models Framework},
  address={Suzhou, China},
  booktitle={Proceedings of the 2025 Conference on Empirical Methods in Natural Language Processing},
  doi={10.18653/v1/2025.emnlp-demos.29},
  pages={418--439},
  publisher={Association for Computational Linguistics},
  topic={nlp},
  url={https://aclanthology.org/2025.emnlp-demos.29},
  year={2025},
  month={11}
}
\end{lstlisting}
\end{tcolorbox}

\vspace{0.8em}
\begin{tcolorbox}[enhanced,
                 frame hidden,
                 boxrule=0pt,
                 borderline west={2pt}{0pt}{Primary},
                 colback=LightBg,
                 sharp corners,
                 breakable,
                 fonttitle=\sffamily\bfseries\large,
                 coltitle=Primary,
                 title=Online Access,
                 attach title to upper={\vspace{0.2em}\par},
                 left=12pt]

\renewcommand{\arraystretch}{1.5}
\begin{tabular}{@{}p{0.25\textwidth}@{}p{0.75\textwidth}@{}}
\textbf{\sffamily Official Publication} & 
\begin{minipage}[t]{0.72\textwidth}
\href{https://aclanthology.org/2025.emnlp-demos.29}{\color{Primary}\url{https://aclanthology.org/2025.emnlp-demos.29}}
\end{minipage}\\

\textbf{\sffamily CiteAssist} & 
\begin{minipage}[t]{0.72\textwidth}
\href{https://citeassist.uni-goettingen.de/preprint/4d2fadc6-f0ae-4623-995a-6fe828bb9aee}{\color{Primary}https://citeassist.uni-goettingen.de/preprint/4d2fadc6-f0ae-4623-995a-6fe828bb9aee}
\end{minipage}\\
\end{tabular}

\end{tcolorbox}

\vspace{0.8em}
\begin{tcolorbox}[enhanced,
                 frame hidden,
                 boxrule=0pt,
                 borderline west={2pt}{0pt}{Primary},
                 colback=LightBg,
                 sharp corners,
                 breakable,
                 fonttitle=\sffamily\bfseries\large,
                 coltitle=Primary,
                 title=Related Papers,
                 attach title to upper={\vspace{0.2em}\par},
                 left=12pt]
\begin{itemize}\itemsep 2pt 
  \item Jonas Becker. \href{https://arxiv.org/abs/2410.22932}{Multi-Agent Large Language Models for Conversational Task-Solving}. 2024. 
  \item Jonas Becker et al. \href{https://arxiv.org/abs/2502.19559}{Stay Focused: Problem Drift in Multi-Agent Debate}. 2025. 
  \item Lars Benedikt Kaesberg et al. \href{https://arxiv.org/abs/2502.19130}{Voting or Consensus? Decision-Making in Multi-Agent Debate}. 2025. 
\end{itemize}
\end{tcolorbox}

\vfill
\begin{tikzpicture}
\draw[Primary!40, line width=0.4pt] (0,0) -- (\textwidth,0);
\end{tikzpicture}
\begin{center}
\small\sffamily\textcolor{TextMuted}{Generated \today}
\end{center}

\end{document}